\documentclass{jltp}

\usepackage{graphicx} % uncomment this line to include the graphicx package

\title{Variational Monte Carlo for Interacting Electrons in Quantum Dots}

\author{Ari Harju}

\address{Laboratory of Physics, Helsinki University of Technology,\\ P.O. Box
1100, FIN-02015 HUT, Finland}

\runninghead{A. Harju}{Variational Monte Carlo for Interacting Electrons in Quantum Dots}

\newcommand{\balpha}{\mbox{\boldmath $\alpha$}}

\begin{document}

\maketitle

\begin{abstract}
  
  We use a variational Monte Carlo algorithm to solve the electronic
  structure of two-dimensional semiconductor quantum dots in external
  magnetic field. We present accurate many-body wave functions for the
  system in various magnetic field regimes. We show the importance of
  symmetry, and demonstrate how it can be used to simplify the
  variational wave functions. We present in detail the algorithm for
  efficient wave function optimization. We also present a Monte Carlo
  -based diagonalization technique to solve the quantum dot problem in
  the strong magnetic field limit where the system is of a
  multiconfiguration nature.

\end{abstract}

%\tableofcontents

\section{INTRODUCTION}

The nanoscale semiconductor systems are technically very promising for
future components of microelectronic devices.  From theoretical point
of view, quantum dot (QD) systems are a valuable source of novel
quantum effects. Many of these result from the fact that the
electron-electron interaction and external magnetic field have greatly
enhanced effects compared to atoms and molecules.  This raises new
challenges for the theoretical methods, and the validity of
approximations in, e.g., mean-field approaches (See
Ref.~\onlinecite{RM} for a review) can be questioned. For this reason,
QD systems serve as perfect test cases to develop electronic structure
methods, with the results applicable to a great variety of physical
problems where mean-field approaches have been used. In addition, many
of the system parameters are tunable, e.g., the electron number in QDs
can be changed one by one and the confinement potential can be varied
by external gates.

The quantum Monte Carlo (QMC) methods are among the most accurate ones
for tackling a problem of interacting quantum particles\cite{QMCRMP}.
Often the simplest QMC method, namely the variational QMC (VMC), is
able to reveal the most important correlation effects. In many quantum
systems, further accuracy in, e.g., energy is needed, and in these
cases methods such as the diffusion QMC (DMC) allow one to obtain more
accurate estimates for various observables. Even DMC is still not
always exact. For fermions, the standard application of the DMC is
variational {\sl with the given nodes}. In finite magnetic fields, the
fixed-node method is generalized to a fixed-phase one, being again
variational\cite{fp}. The ground-state wave function of a bosonic
system (in zero magnetic field) does not have nodes in the ground
state, and DMC has only statistical error in the observables. For
finite magnetic fields, the bosonic problem has a non-trivial phase
structure, and also in that case a fixed-phase strategy has to be
used.

\section{MODEL AND METHODS}

\subsection{Model for Quantum Dots}\label{junk}

The generally used model Hamiltonian of an $N$-electron QD system can
be written as
\begin{equation}
 \mathcal{H}=\sum_{i=1}^N \left\{ \frac{(-\mathrm{i} \hbar \nabla_i
-\frac {e}{c } \mathbf{A})^2}{2 m^*} +
V_{\mathrm{ext}}(\mathbf{r}_i)\right\} + \sum_{i<j}^N
\frac{e^2}{\epsilon|{\mathbf r}_i-{\mathbf r}_j|} +  
g^* \mu_B B S_{z} \ ,
\label{ham}
\end{equation}
where we have used the effective-mass approximation to describe electrons
moving in the $xy$ plane, surrounded by a background material of, e.g., GaAs
with the effective electron mass $m^*=0.067m_e$ and dielectric constant
$\epsilon=12.7$. The magnetic field ${\bf B}$ is included using the symmetric
gauge ${\bf A}=-B(y{\bf u}_x-x{\bf u}_y)/2$. In addition to the Zeeman
coupling to the electron spin (last term above, GaAs value of $g^*$ is around
-0.44), the magnetic field introduces two new terms to the Hamiltonian, namely
a ``squeeze'' term which is
\begin{equation}
\frac{e^2  B^2}{8 m^* c^2} \sum_{i=1}^N (x_i^2+y_i^2) 
=  \frac 12 m^* \left(\frac{\omega_{c}}{2}\right)^2 \sum_{i=1}^N r_i^2 \ ,
\end{equation}
where $\omega_{c}={e { B} \over m^*c}$ is the cyclotron frequency.
The second one is a ``rotation'' term:
\begin{equation}
 \sum_{i=1}^N \frac{e}{2 m^* c} \mathbf{B}\cdot\mathbf{r}_i\times
\frac{\hbar}{\mathrm{i}} \nabla = \frac{\omega_c}{2} L_z \ ,
\end{equation}
where $L_z$ is the total angular momentum operator
($z$-component). These terms are always present, for all shapes of
QDs. For this reason, in all the cases where a finite $B$ is present,
there is a new harmonic term induced in the potential. The rotation
term is a simple one for the cases that have rotational symmetry, as
then the angular momentum is a good quantum number. For systems with
lower symmetry, one needs to calculate the expectation value of the
angular momentum operator. One can see that the rotation term lowers
the energy of states that are rotating (on average) in the correct
direction. This thus breaks the time-reversal symmetry and induces
currents in the system.

We mainly present results for the case of the parabolic confinement,
namely the one with an external potential
\begin{equation}
V_{\mathrm{ext}}(\mathbf{r})=\frac 12 m^* \omega_0^2 \, r^2 \ ,
\end{equation}
where $\omega_0$ gives the strength of the confinement potential. One should
note that it is possible to combine the external potential and the magnetic
confinement as they are both parabolic. The simplest way to do this is to
define a new confinement strength $\omega$ by
$\omega^{2}=\omega_{0}^{2}+{\omega_{c}^{2}/4}$.

There are two obvious choices for the units to be used. The first one
corresponds to the atomic units, but due to the effective mass and the
dielectric constant, they are called the scaled atomic units. These are
obtained by setting $\hbar=m^*=e=\epsilon=1$.
The second set of units commonly used is the ones of a harmonic
oscillator.  This is mainly applicable to the cases where the
confinement potential is parabolic. A natural unit for the energy is
given by the confinement strength $\hbar \omega$, and for the length,
the simplest choice is to measure it in units of
$l=\sqrt{\frac{\hbar}{m^{*}\omega}}$.  The Hamiltonian in harmonic
oscillator units can be written as
\begin{equation}
\mathcal{H} = \sum _{i=1}^N\left ( -\frac12 \nabla_{i}^{2} +
\frac12 r_{i}^{2} \right ) - \frac{\omega_c}{2 \omega} L_{z} +
\frac{g^* \mu_B B}{\hbar\omega} S_{z} + \sum_{i < j}
\frac{C}{r_{ij}} ,
\label{reH}
\end{equation}
where 
$C$ is the Coulomb strength ($C=\sqrt{{\mathrm{ Ha}}/{\hbar\omega}}
{\sqrt{m^*/m_0}}/{\epsilon}=\sqrt{\mathrm{Ha}^*/\hbar\omega} $, where
$\mathrm{Ha}$ is the Hartree $\approx27.2$~eV, and $\mathrm{Ha}^*$ is the unit
of energy in the scaled atomic units).

\subsection{Single-particle States}

\begin{figure}[hbt]
  \includegraphics[width=0.55\columnwidth]{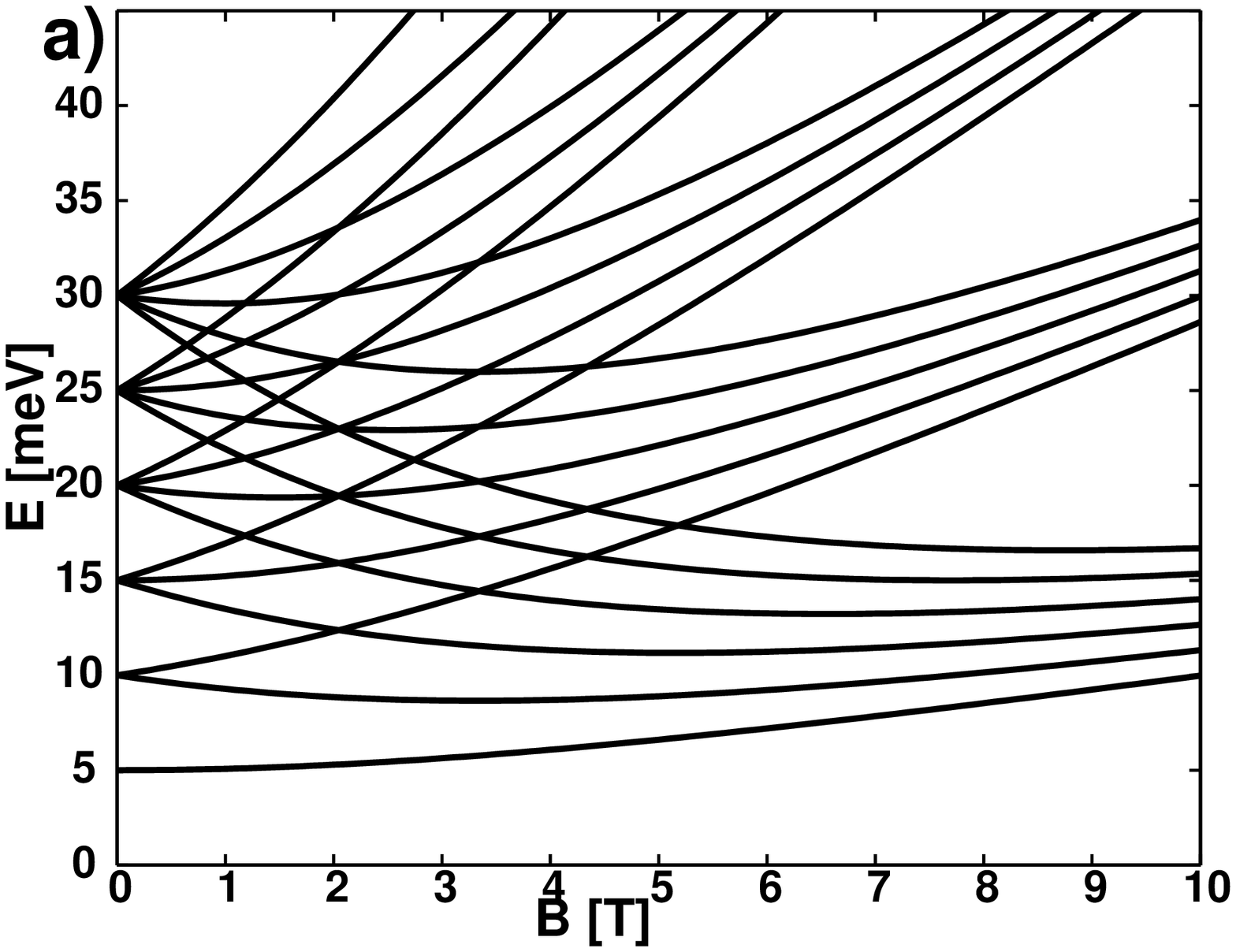}
  \includegraphics[width=0.45\columnwidth]{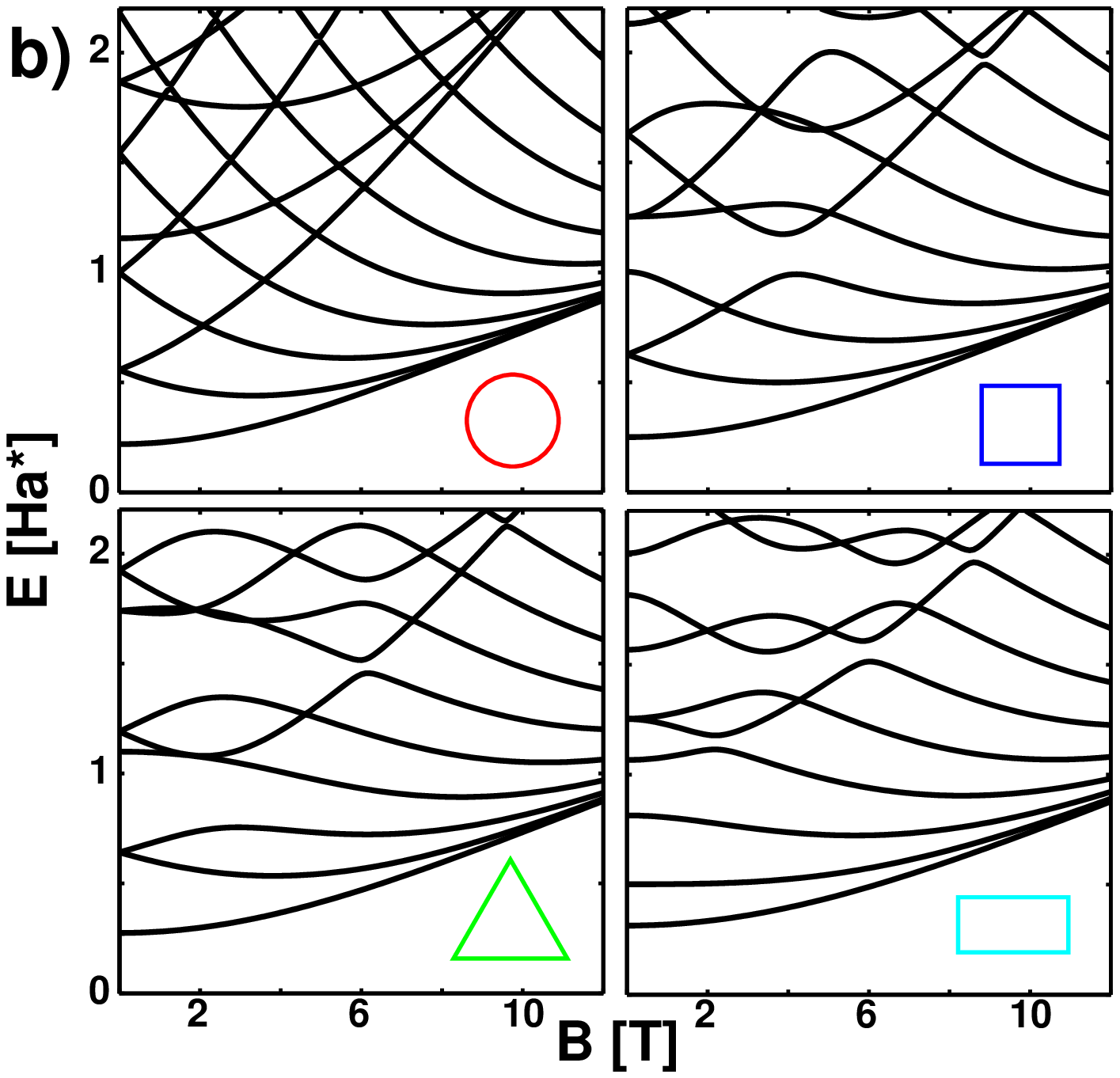}
\caption{(a) Total energy as a function of the magnetic field.  GaAs
  parameter for the effective mass, i.e., $m^*/m_0=0.067$,
  $\hbar\omega_0=5.0$~meV. (b) Same for various hard-wall QDs, see
  Ref.  \onlinecite{rectaB} for more details.}
\label{QD_states_5meV}
\end{figure}

The one-body problem for a parabolic QD is easily solved for an arbitrary
magnetic field\cite{Fock}. The single-particle wave functions (in
harmonic-oscillator units and without the normalization which is irrelevant in
VMC) are
\begin{equation}
\psi_{n, \pm |l|}\propto (x\pm{\mathrm i}y)^{|l|} L_{n'}^{|l|}(r^2) 
\exp\left[-{ r^2}/{2}\right] \ ,
\label{fd}
\end{equation}
where $l$ is the angular momentum quantum number, $n$ is the shell index and
$n'=(n-|l|)/2$, and $L$ is the associated Laguerre polynomial. The two first
polynomials $L$ are
\begin{eqnarray}
L_0^l(x)&=&1\\
L_1^l(x)&=&-x+l+1\ ,
\end{eqnarray}
and any $L$ can be calculated from $L_0^l$ and $L_1^l$ using the recurrence
relation
\begin{equation}
(n+1)L_{n+1}^l(x)=(2n+l+1-x)L_{n}^{l}(x)-(n+l)L_{n-1}^{l}(x) \ .
\label{recurrence}
\end{equation}
The energy of a state with quantum numbers $(n,l)$ is given by
\begin{equation}
E_{n,l} = (n+1) - \frac{\omega_c}{2 \omega}l \ .
\end{equation}
This is plotted in Fig. \ref{QD_states_5meV}(a). One can compare this with
lower-symmetry QDs, shown in Fig. \ref{QD_states_5meV}(b)\cite{rectaB}.
These four have a hard-wall potential of the shape shown in
insets. One can see that the parabolic potential has the highest
symmetry of the spectra shown, with high degeneracy in the $B=0$ limit
as well as a regular set of level-crossings at finite $B$ values. The
lowered symmetry affects the spectra at $B=0$, and the lower the
symmetry is, the more one sees anti-crossings of levels (instead of
crossings).

In all the spectra of Fig. \ref{QD_states_5meV}, one can see that the
lowest states at strong $B$ are bunched together to what corresponds
to the lowest Landau level (LLL). For a parabolic QD, the LLL
functions are of a very simple form, namely
\begin{equation}
\psi_{l}(\mathbf{r})\propto z^{l} \exp\left[-{ r^2}/{2}\right] \ ,
\end{equation}
where $z=x+{\mathrm i} y$, and only the angular momentum quantum
number $l$ is needed to label these states.

\subsection{Non-interacting States}

The non-interacting many-body states can be written as Slater
determinants formed from the single-particle states of
Eq.~(\ref{fd}). As the single-particle states cross as a function of
$B$, the non-interacting ground-state Slater determinants also have
changes in the occupations.  A general trend in the changes is that as
$B$ gets stronger, the occupations are mainly on the LLL. In the
strong-$B$ limit, all electrons occupy a state from the LLL. If we
neglect the Zeeman-coupling to spin, the system has, for an even
number of particles $N$, equal number of up- and down-spin electrons
$N_{\uparrow}$=$N_{\downarrow}$=$N/2$. The non-interacting up-spin
electrons occupy the $N_{\uparrow}$ lowest states, and these have
quantum numbers $l=0,\dots,N_{\uparrow}-1$, and the down-spin electron
have similar occupations. This non-interacting state corresponds to
the filling factor $\nu=2$ integral quantum Hall effect (IQHE)
state. The correspondence to the IQHE state is, of course, only a
qualitative one, as IQHE corresponds more to the thermodynamic
limit. The reason why we now have $\nu=2$ can simply be seen from the
fact that the LLL states are doubly occupied (by both spin types).
The up- and down-spin determinants can be written in a simple fashion,
as they correspond to the Vandermonde determinants. Thus the
unnormalized wave function has the form
\begin{equation}
\Psi(\mathbf{r}_1,\mathbf{r}_2,\dots,\mathbf{r}_N) =
\exp\left[-\sum_{i=1}^N r_i^2/2\right]
\prod_{i_{\uparrow}<j_{\uparrow}}^{N_{\uparrow}} z_{ij}
\prod_{i_{\downarrow}<j_{\downarrow}}^{N_{\downarrow}} z_{ij} \ ,
\label{lllwf}
\end{equation}
where $z_{ij}=z_i-z_j$, and the spin indices are dropped for simplicity.

If all electrons are of the same spin type, one has a state with wave
function
\begin{equation}
\Psi(\mathbf{r}_1,\mathbf{r}_2,\dots,\mathbf{r}_N) =
\exp\left[-\sum_{i=1}^N r_i^2/2\right] \prod_{i<j}^N z_{ij} \ ,
\label{mddwf}
\end{equation}
which now corresponds to $\nu=1$, as each LLL state is occupied by one
electron. This state is called the {\sl maximum-density droplet}
(MDD), for the reason that it has the maximum density possible in the
LLL\cite{macdonald}.

For a more general state, Eq.~(\ref{recurrence}) can be used to
simplify the determinant. To see this, consider states that have the
same angular momentum $l$, but different shell index $n$. One can see
in Eq.~(\ref{recurrence}) that the state with shell index $n+1$ can be
obtained in a simple fashion from the states with the same $l$ but
that have shell indices $n$ and $n-1$. The basic determinant
calculation rule tells that one can remove from a row (or column) a
linear combination of the other rows (or columns). Now in a Slater
determinant, this simplification can be used to simplify
Eq.~(\ref{recurrence}) for the states that have all states with the
same $l$ but lower $n$ occupied. This simplified recurrence reads
\begin{equation}
\tilde L_{n'+1}^l(r^2) = r^2 \tilde L_{n'}^{l}(r^2) \ ,
\label{recurrence2}
\end{equation}
where we have also dropped the normalization and used that in QDs, we have
$x=r^2$. Noting that $L_0^l(x)=1$, one finally gets
\begin{equation}
\tilde L_{n'}^l(r^2) = r^{n-|l|} \ .
\end{equation}
The configurations that have all states with same $l$ but lower $n$ occupied
are called $n$-compact. For these configurations, one can use for the one-body
states the simplified form
\begin{equation}
\psi_{n, \pm |l|}\propto (x\pm{\mathrm i}y)^{|l|} r^{n-|l|}
 \exp[-{r^2}/{2}] \ .
\label{simplewf}
\end{equation}
One should note that all the lowest-energy states of a non-interacting
$N$-electron QD are $n$-compact. In addition, there is no obvious
reason why interactions would favor non-$n$-compact states.

\subsection{Separation of the Center-of-Mass Motion}

For a parabolic QD, the Hamiltonian of Eq.~(\ref{reH}) has an
important simplifying property; the center-of-mass (CM) and the
relative motion of electrons decouple. To see this, one has to make
use of the identity
\begin{equation}
\sum_{i=1}^N r_i^2 = N r_{\mathrm{cm}}^2+\frac{1}{N} \sum_{i<j}^N r_{ij}^2 \ ,
\end{equation}
where $\mathbf{r}_{\mathrm{cm}}$ is the coordinate for the CM.

The separation of the CM and relative motion has an interesting
consequence, namely that if the wave function for the non-interacting
state is an eigenstate of the CM motion, then turning on the
electron-electron interaction can change only the relative motion part
of the wave function. This means that the many-body wave function can
be written as a product of the CM and relative wave functions as
\begin{equation}
\Psi = \Psi_{\mathrm{cm}}(\mathbf{r}_{\mathrm{cm}}) 
\Phi(\{\mathbf{r}_{ij}\}_{i<j}) \ ,
\end{equation}
where the $\Phi$-part contains all non-trivial effects of
interactions, and the CM-part is a solution for a harmonic oscillator,
see, e.g., Ref.~\onlinecite{Bolton} for more details. To see how this
affects the topology of the wave function, one should note that the
exponential factor is the same in all states of Eq.~(\ref{fd}), and it
can be taken out from any determinant made of these as
\begin{equation}
\exp\left[-\frac12 \sum_{i=1}^N r_i^2\right] = \exp\left[-\frac12 N
r_{\mathrm{cm}}^2 \right]\exp\left[-\frac{1}{2N} \sum_{i<j}^N
r_{ij}^2\right] \ .
\end{equation}
Now if one would like to include a variational parameter $\alpha$ to
the exponential as $\exp[- \alpha r_i^2/2]$, one would directly see from
the argument above that the optimal value for $\alpha$ should be
one. Other choices would raise the energy because the CM-part would
not be exact.  In addition, even if varying $\alpha$ would lower the
interaction energy, this kind of effect should be directly put in to
the wave function of relative coordinates, for example using two-body
correlation factors.  For this reason there is no similar screening of
the external potential as in the case of atoms. If one, however, would
apply a Hartree-Fock approach for the problem, one certainly lowers
the energy by adjusting the single-particle states. This is because
one cannot change the relative-motion part of the Hartree-Fock wave
function as it is, by definition, only a single Slater determinant.

We can thus have the following important conclusion: If the relative
motion part of the many-body wave function is treated in a reasonable
accuracy, the single-particle states in the determinant part of the
Slater-Jastrow wave function (see Section~\ref{WFs}) can be taken to
be the non-interacting ones. In many VMC studies, one uses
single-particle states from a mean-field theory in the construction of
the Slater determinants. This is not, however, the most accurate
strategy, based on the arguments presented above\cite{cyrus}.

The separation has also a second effect, namely that the far-infrared
absorption spectra of parabolic QDs are trivial, and do not depend on
electron number or the interaction between electrons. For lower
symmetries of the confinement potential, interesting spectra are
obtained\cite{FIR,fir2}.

\subsection{Wave Functions for Quantum Dots}\label{WFs}

The most commonly used VMC wave function is the Slater-Jastrow one of
the type:
\begin{equation}
\Psi = D_{\uparrow} D_{\downarrow} \prod_{i<j}^N J(r_{ij}) \ ,
\label{wf}
\end{equation}
where the two first factors are Slater determinants for the two spin
types, and $J$ is a Jastrow two-body correlation factor. We neglect
here the three-body and higher correlations. This form of a wave
function has shown to be very accurate in many cases\cite{QMCRMP}.

One can easily generalize the Jastrow part to also contain
higher-order correlations. This might not be very important in the
present case, as we are studying a two-dimensional system. Basically,
by lowering the dimensionality of the problem, one enhances the
correlation effects, because particles have less degrees of freedom to
avoid each other. On the other hand, the lowered dimensionality makes
it more difficult for more than two particles to get close to each
other, and in this way the relative importance of the correlations
beyond the two-particle level gets smaller as the dimensionality is
lowered\cite{thanks1}.

For the two-body Jastrow factor we use here a simple form of
\begin{equation}
J(r)=\exp\left({\frac{C r}{a+b r}}\right) \ ,
\label{Jsimple}
\end{equation}
where $a$ is fixed by the cusp condition to be 3 for a pair of equal
spins and 1 for opposite ones, and $b$ is a parameter, different for
both spin-pair possibilities. $C$ is the scaled Coulomb strength.  For
a $d$ dimensional Coulombic system, the cusp condition can actually be
found to be
\begin{equation}
\frac{J'(r)}{J(r)}=\frac{C}{d-1+2 l} \ ,
\end{equation}
where $l$ is the relative angular momentum between particles. The
actual value of $l$ depends on the wave function form, but for
electrons with opposite spins $l \ge 0$ and for parallel spins $l \ge
1$.

In the simplest form of the determinant in Eq.~(\ref{wf}), the
single-particle wave functions depend only on the coordinates of one
electron. However, in some complicated cases the single-particle
states must depend on all particles. This can be the case for some QD
states, too. If one would recast this type of determinant to a basic
one with simple single-particle states, more than one determinant is
needed.

The basic reason why the Slater-Jastrow wave function is accurate in
many quantum systems is that there is one most important configuration
in an exact expansion of the wave function in terms of configurations.
In addition, the next configuration (of the same symmetry) is in the
successful cases clearly higher in energy. One can then view the
addition of the Jastrow factor as a minor correction to a
single-determinantal wave function.

This reasoning might, however, be too trivial for QD's. As shown in
Section \ref{twoe} for the case of two electrons, the Slater-Jastrow
form of Eq.~(\ref{wf}) is actually exact for any interaction
strength. This means that even in the limit $C\to \infty$ where the
overlap of a single determinant wave function with the exact one goes
to zero, the Slater-Jastrow combination is exact. A similar finding
can be seen in a six-electron QD at $C\to \infty$, see Section
\ref{zero} below.

The arguments given above work reasonably well from the zero magnetic
field case up to rather strong magnetic fields. However, when the
magnetic field is so strong that the system is in a fractional quantum
Hall effect regime, things change dramatically. The interesting and
rich physics of this regime is namely given by the fact that there is
a high degeneracy of single-particle states in the lowest Landau
level. There are cases when the wave function in this regime can be
written as a Slater-Jastrow one, and the Laughlin states are
interesting examples of this\cite{Laughlin_13}. But there are also
cases when such a simplification might not work well.

\subsection{Variational Monte Carlo}

A brief review of the variational Monte Carlo (VMC) method can be
found in Ref.~\onlinecite{QMCRMP}. In our implementations, we
typically perform VMC with around ten walkers at the same time. There
are two basic reasons for this, namely that as these walkers are
independent, we can do a rigorous error estimate over them. In
addition, these are needed for the wave function optimization
presented below.

If our $N_W$ walkers have estimates $\{a_i\}_{i=1}^{N_W}$ for an
observable $O$, the best estimate for $O$ is the mean of $a_i$'s. The
error estimate can be obtained from the law of large numbers, and it
states that we can approximate the error in our estimate as the
standard deviation of the set $\{a_i\}_{i=1}^{N_W}$ divided by the
square root of $N_W$. If one would have only one walker, one should
perform by some means an analysis of the autocorrelation length for
the observable in question to get a rigorous error estimate.

Lin {\it et al.}\cite{anad} have shown that in the case of real wave
functions and energy minimization, the derivative of the energy $E$
with respect to a variational parameter $\alpha_i$ is simply
\begin{eqnarray}
\frac{\partial E}{\partial \alpha_i} &=& 
\frac{\partial }{\partial \alpha_i} \frac{\int \Psi \mathcal{H} \Psi}{\int \Psi^2}
\nonumber\\
&=&
\frac{\int \Psi' \, \mathcal{H} \, \Psi}{\int \Psi^2}
+\frac{\int \Psi \, \mathcal{H} \, \Psi'}{\int \Psi^2} 
-2 \frac{\int \Psi \, \mathcal{H} \, \Psi }{\int \Psi^2}
 \frac{\int \Psi' \, \Psi}{\int \Psi^2}
\nonumber\\
&=& 
2\left\langle E_L 
\frac{\Psi'}{\Psi} \right\rangle -2\left\langle E_L\right\rangle 
\left\langle \frac{\Psi'}{\Psi} \right\rangle \ ,
\label{anaa}
\end{eqnarray}
where $\Psi'={\partial \Psi}/{\partial \alpha_i}$,
$E_L=\frac{\mathcal{H}\Psi}{\Psi}$ is the local energy and the average
$\langle \dots \rangle$ is over a set of configurations generated
with, e.g., the Metropolis algorithm\cite{anad}.

Some possibilities for optimizing the parameters are presented in
Ref.~\onlinecite{QMCRMP}. These do not benefit from the simple formula
for the gradient of the parameters given above. Especially for this
reason, we think that the optimization can be done in a more efficient
fashion than is explained in Ref.~\onlinecite{QMCRMP}. One possibility
for this is the stochastic gradient approximation (SGA)
method\cite{sga}, which is an optimization method tailored for
functions with statistical uncertainty. The SGA algorithm in the
original formulation did not use the simple gradient formula.

Given a set of $N_W$ walkers, one can calculate an approximation for
the gradient of the optimized parameters with respect to the cost
function by Eq.~(\ref{anaa}). To minimize the cost function, one
should move in the direction of the negative gradient. Unfortunately,
the step length needed is not trivial to estimate. In SGA, the step is
adjusted dynamically by a parameter $\gamma$.  So at the $i$'th
optimization step, the parameters are changed by
\begin{equation}
\balpha_{i+1}=\balpha_i-\gamma_i \nabla_{\balpha} F_C \ ,
\end{equation}
where $F_C$ is the cost function. The step length $\gamma_i$ is
changed in such a fashion that it monitors the changes in sign of the
gradient. There is a simple reason for this: if the sign of the
gradient remains the same over the steps, one is approaching the
minimum of the cost function and there is no reason to slow down
(actually, one could increase the step length). If the sign of the
gradient oscillates, one most probably makes too long steps in the
optimization and the minimum is somewhere close by and the
optimization jumps each time over it to the other side of the
minimum. Thus it would be better to make $\gamma$ smaller. Of course,
our gradient can change sign simply due to the statistical noise in
the gradient. In actual codes, each parameter has an integer counter
that counts the times the sign of the gradient has changed. The value
of this integer $j$ at step $i$ then defines the value for
$\gamma_i=\gamma_0 \ j^{-0.6}$, where $\gamma_0$ is an adjustable
parameter to give an overall length scale for the optimization
steps. The choice of power $-0.6$ is not arbitrary, see
Ref.~\onlinecite{sga} for limits for it. In actual calculations, one
can start with a small value of $\gamma_0$ and make it larger every
time the gradient keeps a same sign. This can also be done in an
efficient fashion by generalizing the integers $j$ to a real numbers
that start from one but then are, e.g., halved at each step where the
sign of the gradient does not change. After a first change of sign is
obtained, one can then switch to damping (like $j=j+1$) as one has
then reached the surrounding of the minimum being searched.

\begin{figure}[hbt] 
\begin{center}
 \includegraphics[width=0.99\columnwidth]{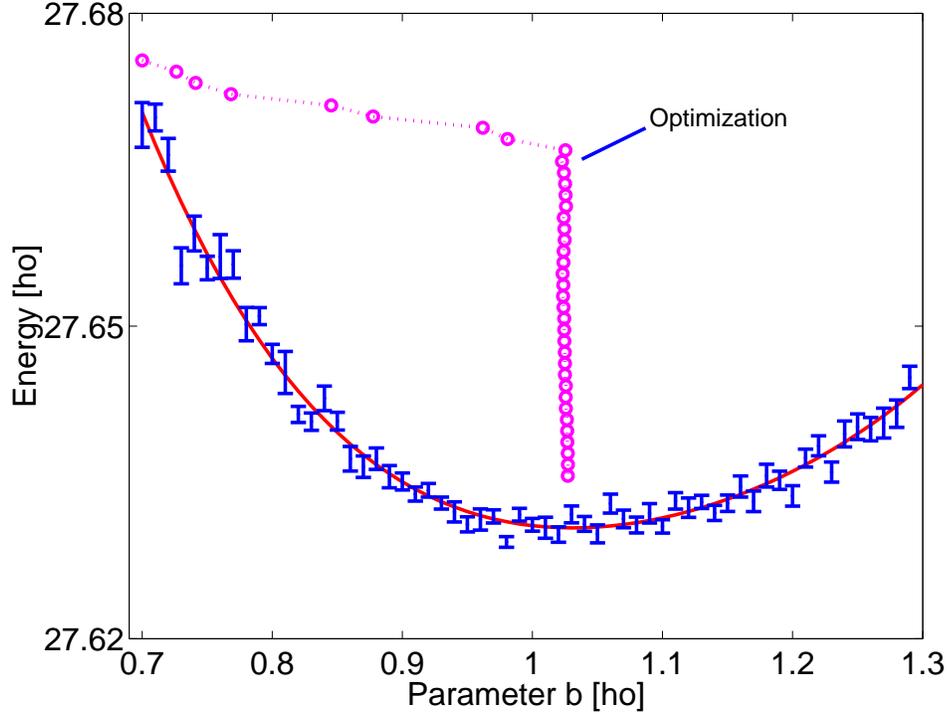}
\caption{Total energy as a function of the Jastrow parameter
$b$. Solid line is a polynomial fit to energies, shown to guide the
eye.  Circles present the SGA parameter optimization path. Each energy
value is from a 10 second run, and all the optimization steps shown
take 0.1 seconds.}
\label{SGA1}
\end{center}
\end{figure}
\begin{figure}[hbt] 
\begin{center}
 \includegraphics[width=0.99\columnwidth]{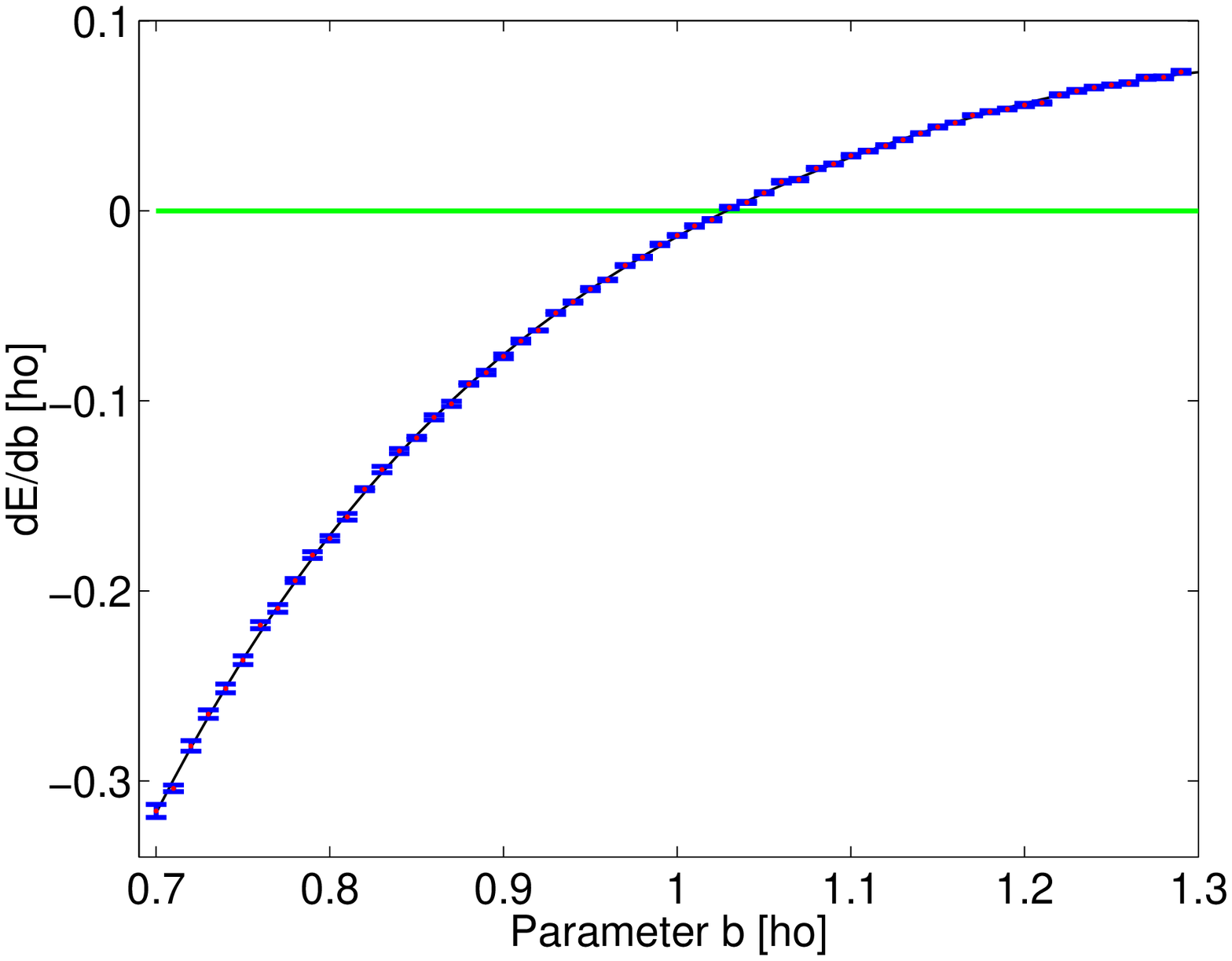}
\caption{Derivative of the total energy with respect to Jastrow
parameter $b$ as a function of $b$. Each derivative from a 10
second run like in Fig.~\ref{SGA1}. It is interesting to note
that the gradient has much smaller fluctuation than the energy.}
\label{SGA2}
\end{center}
\end{figure}
To show the performance of the SGA scheme in actual simulations, a
six-electron MDD case is studied. The wave function is thus a product
of the wave function of Eq.~(\ref{mddwf}) multiplied by a product of
Jastrow factors in Eq.~(\ref{Jsimple}). We have set $C=1$, and the
results are reported in units of the harmonic oscillator. In this
case, we only need to optimize one parameter, the $b$ in Jastrow of
Eq.~(\ref{Jsimple}). First, we have performed 60 short simulations of
10 seconds each for $b$ values between 0.7 and 1.3. The resulting
energies and the error estimates are shown in Fig.~\ref{SGA1}. One can
see that the energy has a minimum value around $b\approx 1$, the
actual position being unclear due to the statistical noise. In
Fig.~\ref{SGA1} we also plot the parameter values obtained by the SGA
optimization that has started from $b=0.7$ and ends around $b\approx
1.03$.  We have used $N_W=10$ walkers in optimization. One can see
that it takes less than ten steps to reach the optimum value. It is
clear from these numbers that the optimization is very fast. The time
needed for all the optimization steps shown is only 1/100 of each of
the independent simulations done, that is, only 0.1 seconds.

\begin{figure}[hbt] 
\begin{center}
 \includegraphics[width=0.9\columnwidth]{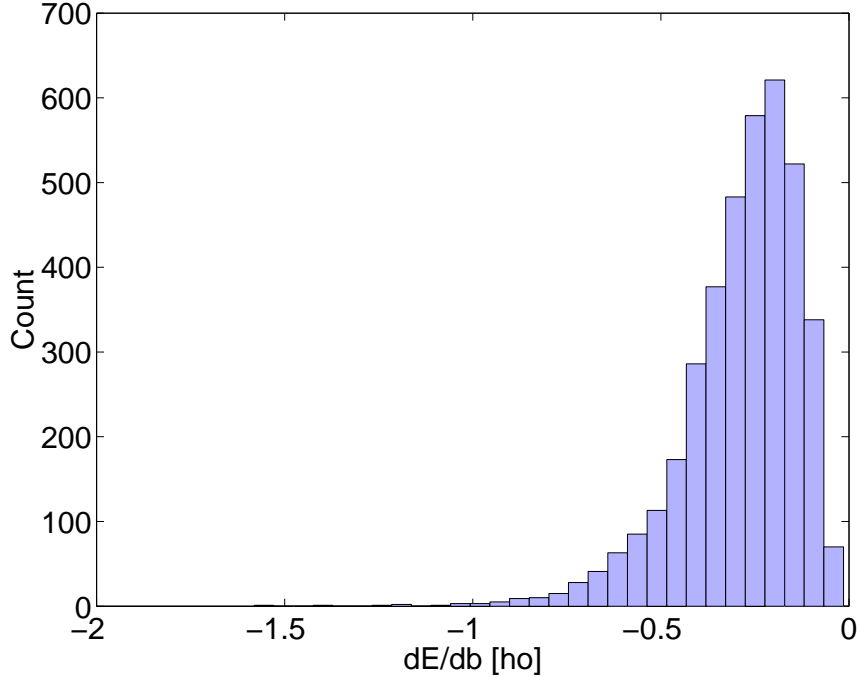}
\caption{The histogram of the gradient values for the case of
$b=0.7$. All the gradient values have the same sign.}
\label{SGA3}
\end{center}
\end{figure}

One reason why the optimization is so fast can be seen in
Fig.~\ref{SGA2}, where the derivative of the energy is plotted. One
can namely see that the statistical noise in the gradient is much
smaller than in the energy. The data of Fig.~\ref{SGA2} also shows
that the parameter value of $b\approx 1.03$, found by the SGA
algorithm, is a very accurate one. In addition to having smaller
fluctuation than energy, also the underlying statistical distribution
of the analytic gradient is in many cases very good for the
optimization.  In Fig.~\ref{SGA3}, the histogram of the gradient
values is shown for the case of $b=0.7$ for a set of around 4000
evaluations of the gradient. An interesting feature is that all the
gradient values in this set have the same sign. This is very good for
the SGA scheme where the actual absolute value of the gradient is less
important than the sign of the gradient.
One should note that closer to the minimum, the gradient distribution
of the present case changes to a rather symmetric one.

One can actually show that in a case of a one-dimensional harmonic
oscillator, defined by the Hamiltonian
\begin{equation}
\mathcal{H}=-\frac 12 \frac{d^2}{dx^2} + \frac 12 x^2 \ ,
\end{equation}
and using a variational wave function of
\begin{equation}
\psi(x)=\exp{(-\alpha x^2)} \ ,
\end{equation}
the analytic gradient formula gives a rather interesting optimization
problem where the distribution of the gradient is such that no matter
how small sample one has in Eq.~(\ref{anaa}), the sign of the energy
gradient is always correct. The gradient is found to be
\begin{equation}
\frac{d E}{d \alpha} = \left(1-\frac{1}{4 \alpha^2}\right) \sigma^2 \ ,
\end{equation}
where $\sigma^2$ is a variance of a function over points sampled from
$\psi^2$, and thus positive.  One thus ends up with a stochastic
optimization problem where the value of the optimized function and its
derivative with respect to parameter is know only with a statistically
limited accuracy, but the sign of the gradient has {\sl not}
statistical noise and can be found exactly by very limited statistics.
Interestingly, similar optimization problem is found for the Hydrogen
atom with wave function $\psi(r)=\exp(-\alpha r)$. It might be that
the special form of the optimization problem is related to the
underlying Gaussian and Poisson distributions and their width which is
proportional to optimized parameter.

\section{RESULTS}

The results are split into three parts, those for zero magnetic field,
moderate magnetic field, and high magnetic field. The splitting is, of
course, not unique, but we feel it is justifiable. By a moderate
magnetic field we mean magnetic field values where the system, in the
thermodynamic limit, would be in an IQHE state. The strong magnetic
field then corresponds to a fractional quantum Hall effect (FQHE)
regime. The logic behind this division is that the physics in these
regimes is rather different.  For example, the physics of IQHE can
still be described, in a reasonable fashion, by arguments on a
single-particle level. However, when the system is in the FQHE regime,
the electron-electron interaction dominates and all single-particle
levels are nearly degenerate. This difference is then reflected in the
nature of the many-body wave functions: In the IQHE regime, a single
configuration is in most cases adequate, but for FQHE-states several
configurations are needed. One would expect that the mean-field
theories are still reasonably accurate in the IQHE regime, but for the
FQHE cases, a real many-body theory is needed. The reason for
splitting the zero magnetic field from the moderate one is that even a
small but finite magnetic field changes the physics in an important
fashion. For example, the time-reversal symmetry is broken as soon as
finite magnetic field is included. On the other hand, the transitions
in finite quantum systems are in general either smooth or of the
level-crossing-type. For this reason the split introduced here might
not always be clear-cut. Before presenting real many-body results, the
results of the simplest non-trivial case, namely the two-electron one,
are presented. As will be shown below, it allows one to understand
many properties of the results for larger particle numbers.

\subsection{Two Electrons}\label{twoe}

The problem of two electrons in a parabolic confinement, subject to
external magnetic field, is in principle four-dimensional. However,
after separating the center-of-mass and relative motion and using the
rotational symmetry, the problem is reduced to a one-dimensional one.

As a special case, the one where the relative interaction strength
$C=1$ in Eq.~(\ref{reH}) and $B=0$, the (unnormalized) solution can be
written as
\begin{equation}
\Psi(\mathbf{r}_1,\mathbf{r}_2) = (1+r_{12}) \exp[-(r_1^2+r_2^2)/2] \ ,
\end{equation}
with energy $E=3$.

This state has a simple generalization for finite $B$ and any interaction
strength, namely
\begin{equation}
\Psi ({\bf r}_{1}, {\bf r}_{2}) = z_{12}^{m} \ J(r_{12}) \ 
\exp\left [-(r_{1}^{2}+r_{2}^{2})/{2}\right ] ,
\label{2e}
\end{equation}
where $m$ is the relative angular momentum of the electrons.  Varying
the interaction strength only changes the $J$-part of the wave
function. It turns out that a simple Jastrow factor of the type
\begin{equation}
J({r})=\exp\left [\frac{
{C r}}{(2|m|+1)(1+b{r})}\right ]
\label{Jastrow}
\end{equation}
is a rather accurate approximation for the true solution\cite{twoe}.

\begin{figure}[hbt] 
\begin{center}
 \includegraphics[width=0.79\columnwidth]{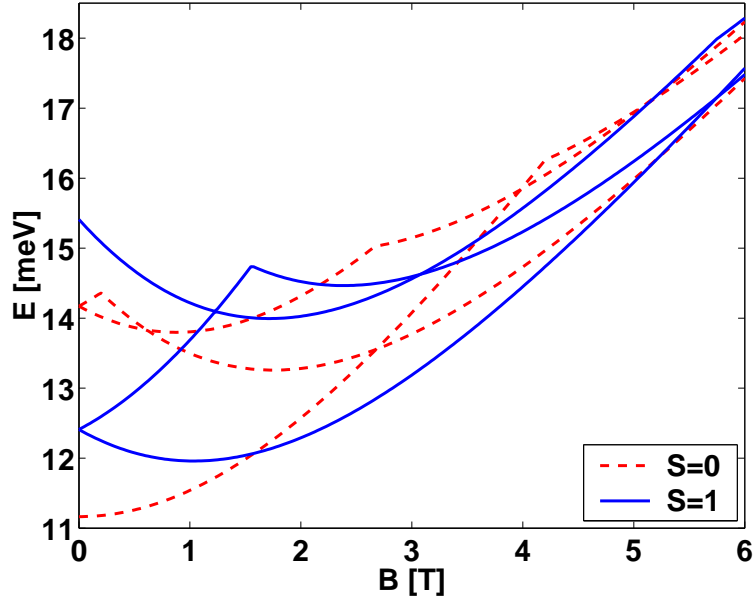}
\caption{Energy for a two-electron GaAs QD. Confinement strength is
3~meV. Two crossings in the ground state are seen.}
\label{Meri}
\end{center}
\end{figure}
In addition to changing the relative-coordinate function $J$ above,
the interaction has a second, even more interesting effect. Namely,
when the magnetic field is made stronger, the interaction favors a
state that has a higher angular momentum than the $B=0$ one with
$m=0$, see Fig.~\ref{Meri}. The reason for this is that the electrons
are moving further apart as $m$ is made larger. The states with even
$m$ correspond to total spin $S=0$, and the odd ones to $S=1$. Now,
without Zeeman term, the spin would oscillate between zero and one,
but if the Zeeman interaction is included, it lowers the energy of the
triplet state, and the strong $B$ ground state is found to be $S=1$.

\begin{figure}[th]
\includegraphics[width=0.49\columnwidth]{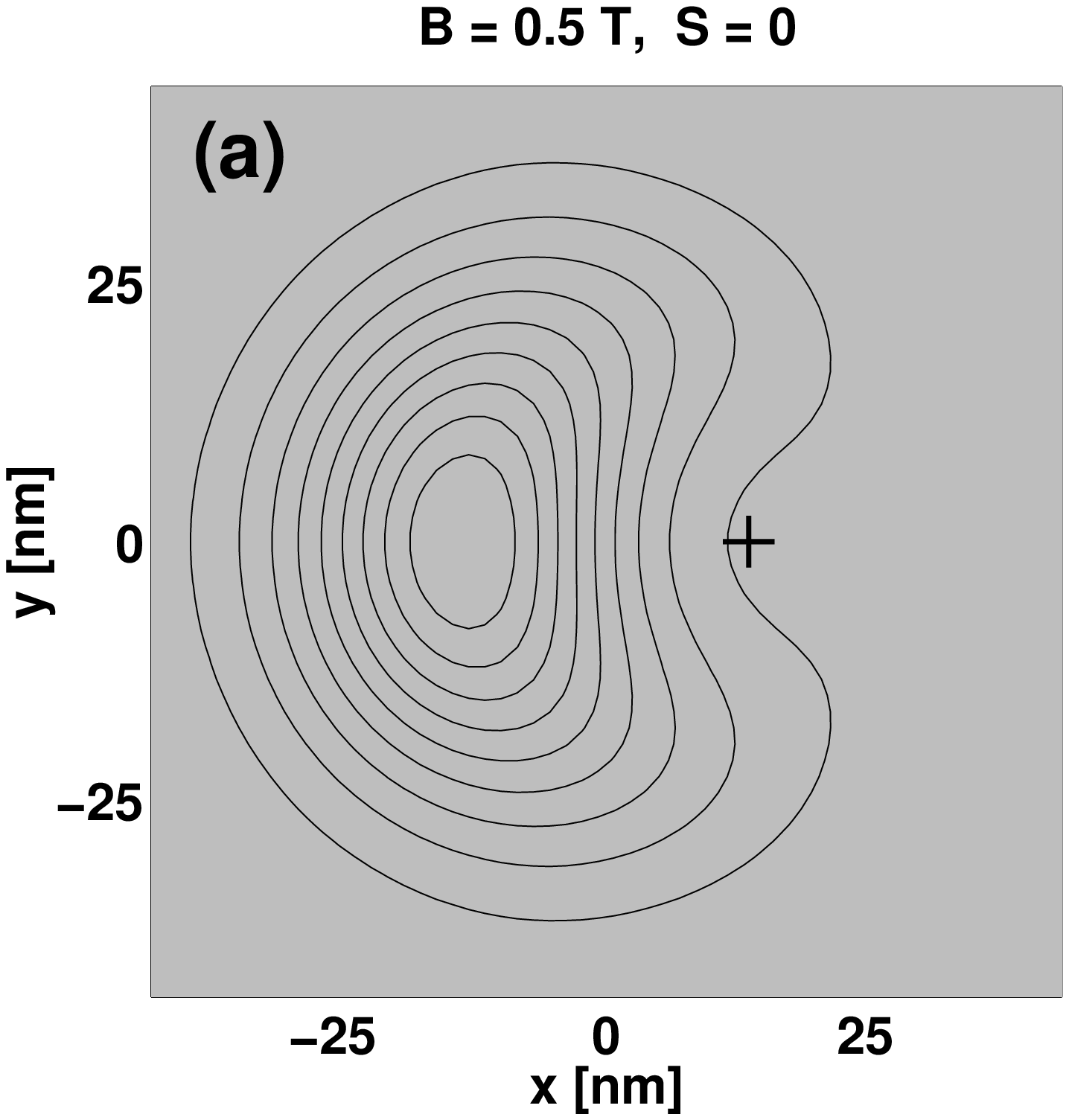}
\includegraphics[width=0.49\columnwidth]{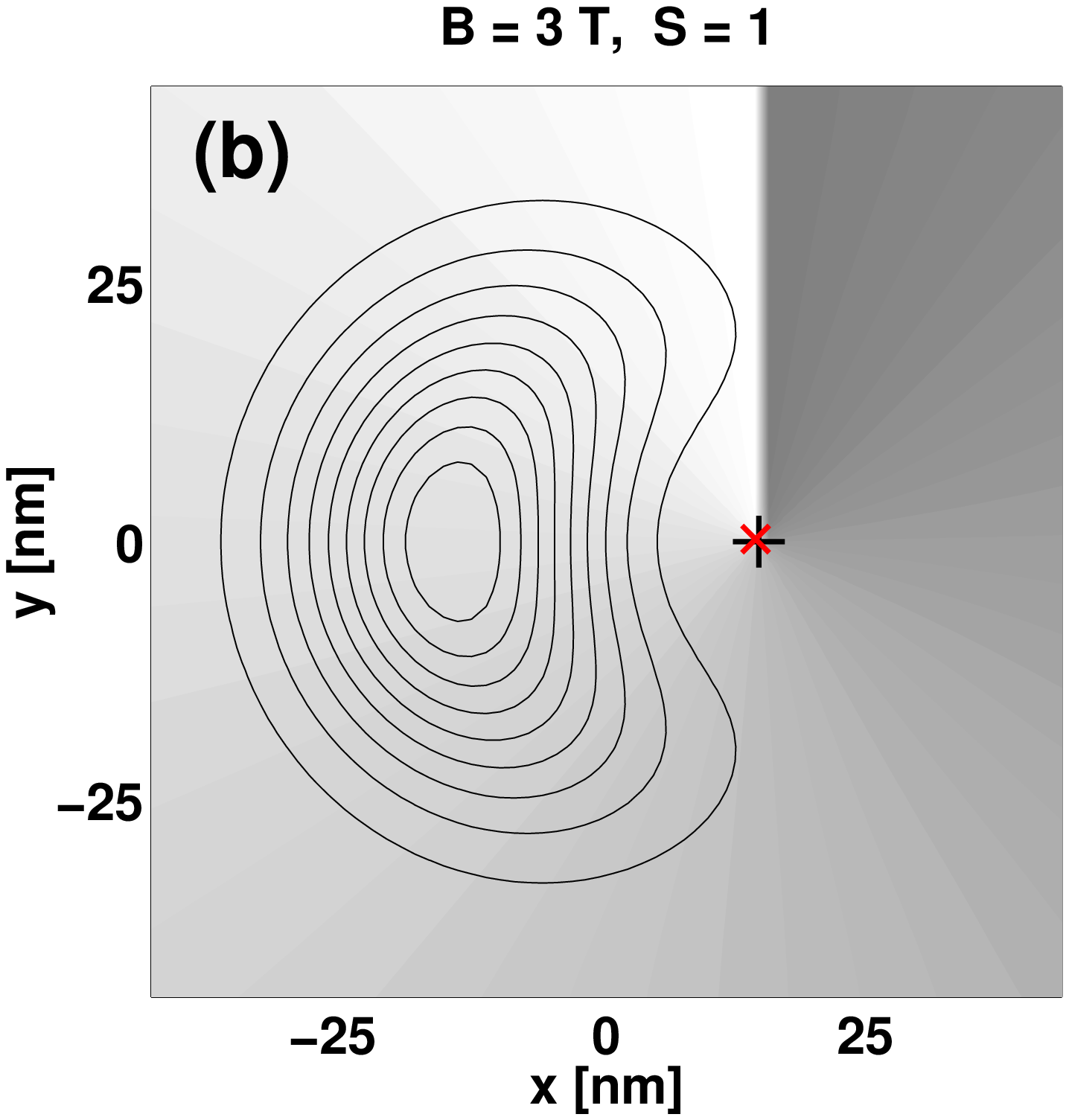}
\caption{Conditional density and phase for the two-electron QD of
Fig.~\ref{Meri}. One electron is fixed at ``$+$'', and the conditional
electron density is presented by contours and the phase by the
gray-scale.  The phase changes from $\pi$ to $-\pi$ on the lines where
shadowing changes from the darkest gray to white. (a) corresponds to
the $S=0$ state at a weak magnetic field, and (b) to the $S=1$ state
with one vortex, marked with ``$\times$'', on top of the fixed
electron.}
\label{2evortices}
\end{figure}

One can analyze the phase structure of the wave function by fixing one
electron and studying the phase as the other electron is
moved\cite{vortex,vortex2,vortex3}. One can see from Eq.~(\ref{2e})
that the phase of the wave function changes by $2 \pi m$ as our probe
electron circulates the fixed one. In addition, the conditional
density is zero at this point. One can identify this zero as a vortex
(with winding number $m$)\cite{vortex}, which actually corresponds to
$m$ flux quanta on top of the fixed electron. In this way, the
magnetic field is used to keep the two electrons further apart.  See
Fig.~\ref{2evortices}.

If one lowers the symmetry of the confinement potential, the wave function is
not as simple anymore. However, some similarities can still be found in the
phase structure of the wave function and in the behavior of the total
spin\cite{qdm}.

\subsection{Zero Magnetic Field}\label{zero}

The most interesting limit in zero magnetic field is the one where the
confinement is extremely weak. In this limit, the interactions
dominate in the energy, and the system approaches the classical
limit\cite{CL}. Before the classical limit, one sees {\sl Wigner
molecule} formation. In this regime, the electrons start to localize
(in relative coordinates), so that if one fixes all but one of the
electrons and studies the conditional density for the remaining one,
the density is seen to get more and more peaked as the system
approaches the classical limit\cite{Wigner}.

\begin{figure}[hbt]
\begin{center}
  \includegraphics[width=0.49\columnwidth]{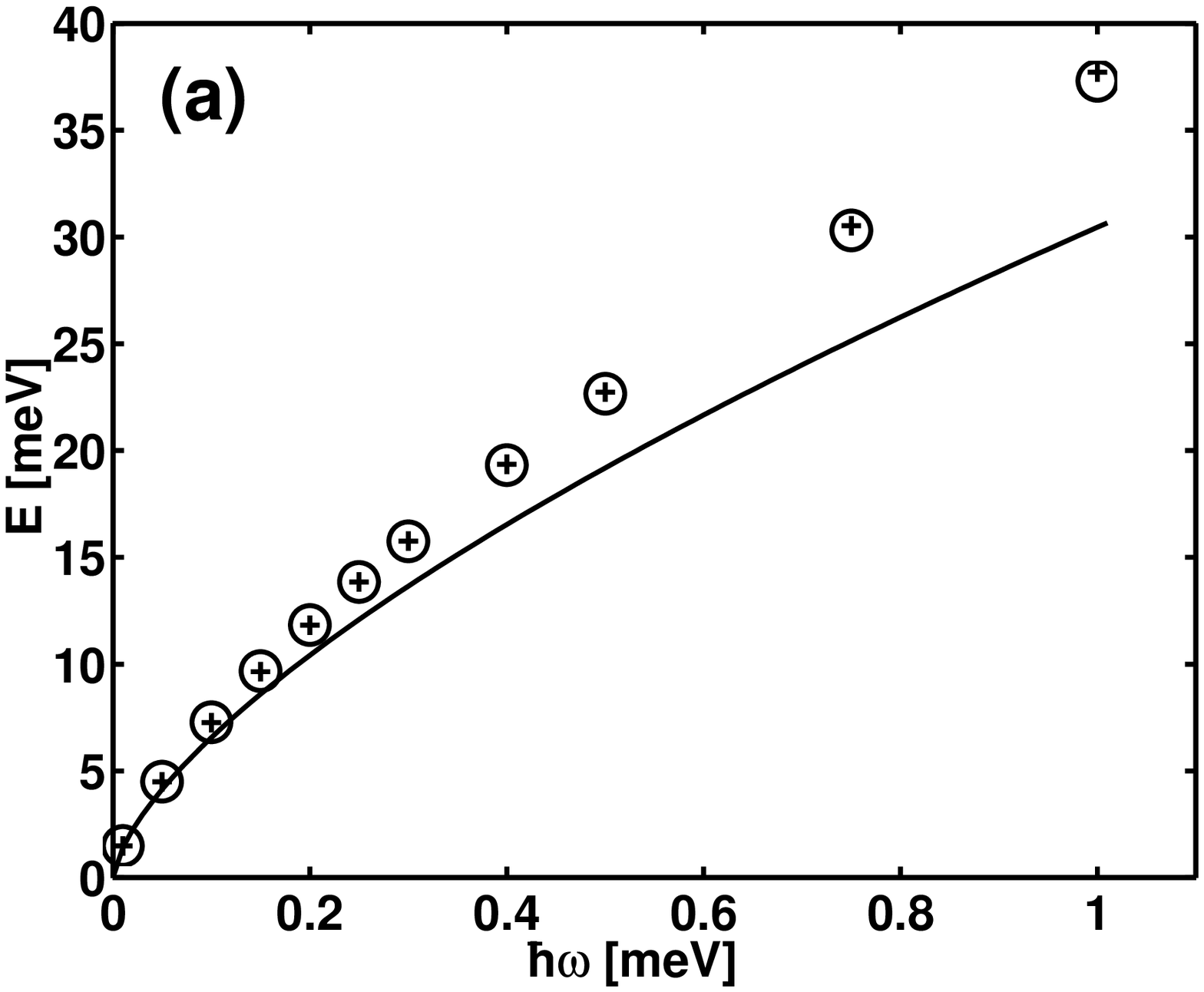}
  \includegraphics[width=0.49\columnwidth]{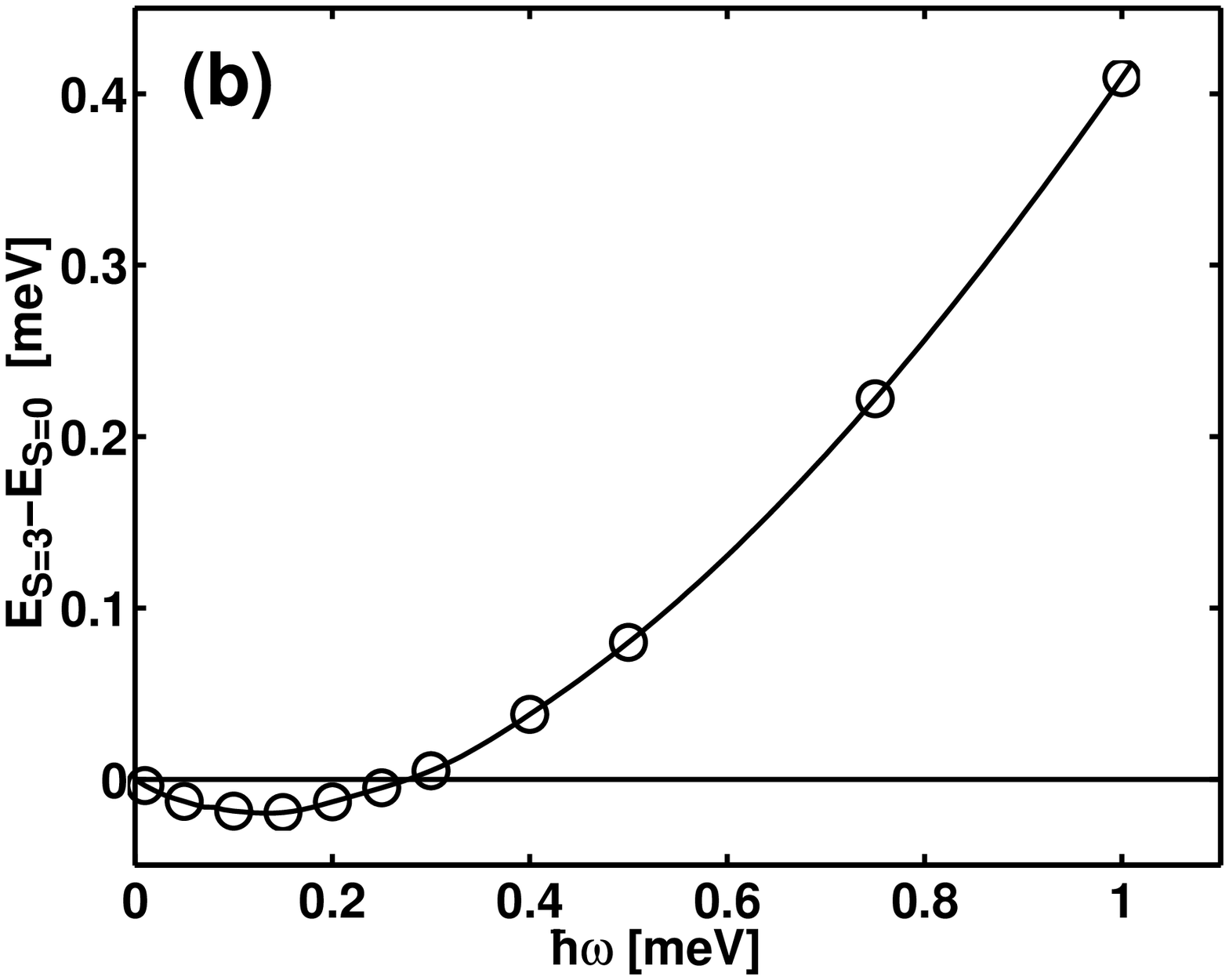}
\end{center}
\caption{(a) Total energy for spin states $S=3$ (marked with pluses)
and $S=0$ (circles) as a function of $\hbar \omega$. The line presents
the classical energy. (b) Energy difference between the spin states
$S=3$ and $S=0$ as a function of $\hbar \omega$. The line is to guide
the eye.}
\label{WC0}
\end{figure}

The energies obtained with a single-determinant-Jastrow wave function
are extremely accurate in the Wigner-molecule limit\cite{Wigner}. The
energies for spin polarized and unpolarized cases are shown in
Fig.~\ref{WC0}(a). For the six-electron QD, we found a spin
polarization in the system before the Wigner-molecule limit, see
Fig.~\ref{WC0}(b). One should note that the exact diagonalization
results are less accurate than the VMC ones, and the results obtained
with a path-integral Monte Carlo have around two orders of magnitude
larger statistical uncertainty and are higher in energy, see
Ref.~\onlinecite{Wigner} for details.
In finite systems like this, the
shell structure still plays a role. For this reason, a six-electron QD is a
good candidate for more general conclusions, as both unpolarized and fully
spin-polarized configurations correspond to closed shells. For open shells,
the wave function might not always be as simple as for the closed-shell
cases\cite{Wigner}.

\begin{figure}[htb]
\begin{center}
  \includegraphics[width=0.425\columnwidth]{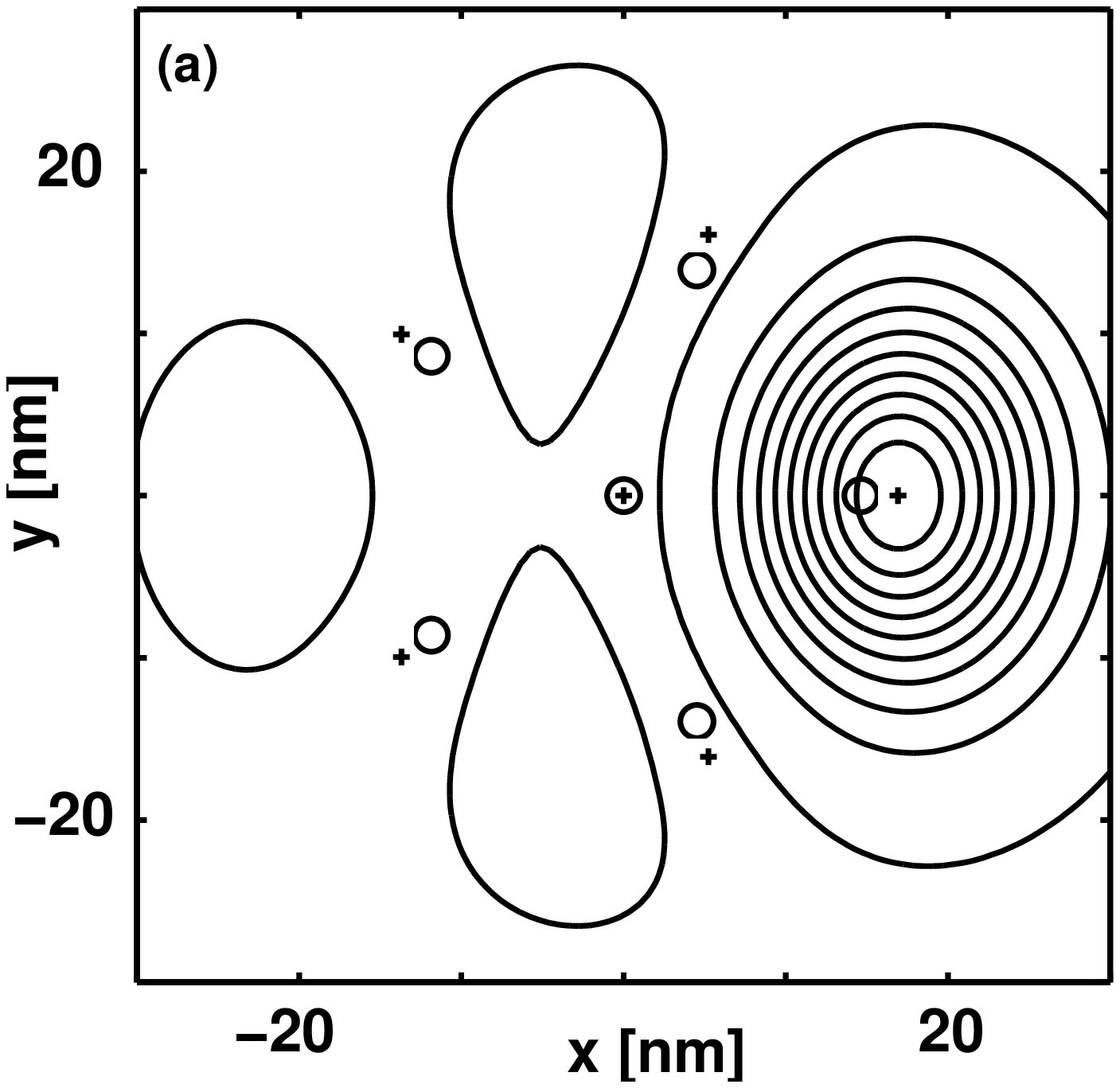}
  \includegraphics[width=0.425\columnwidth]{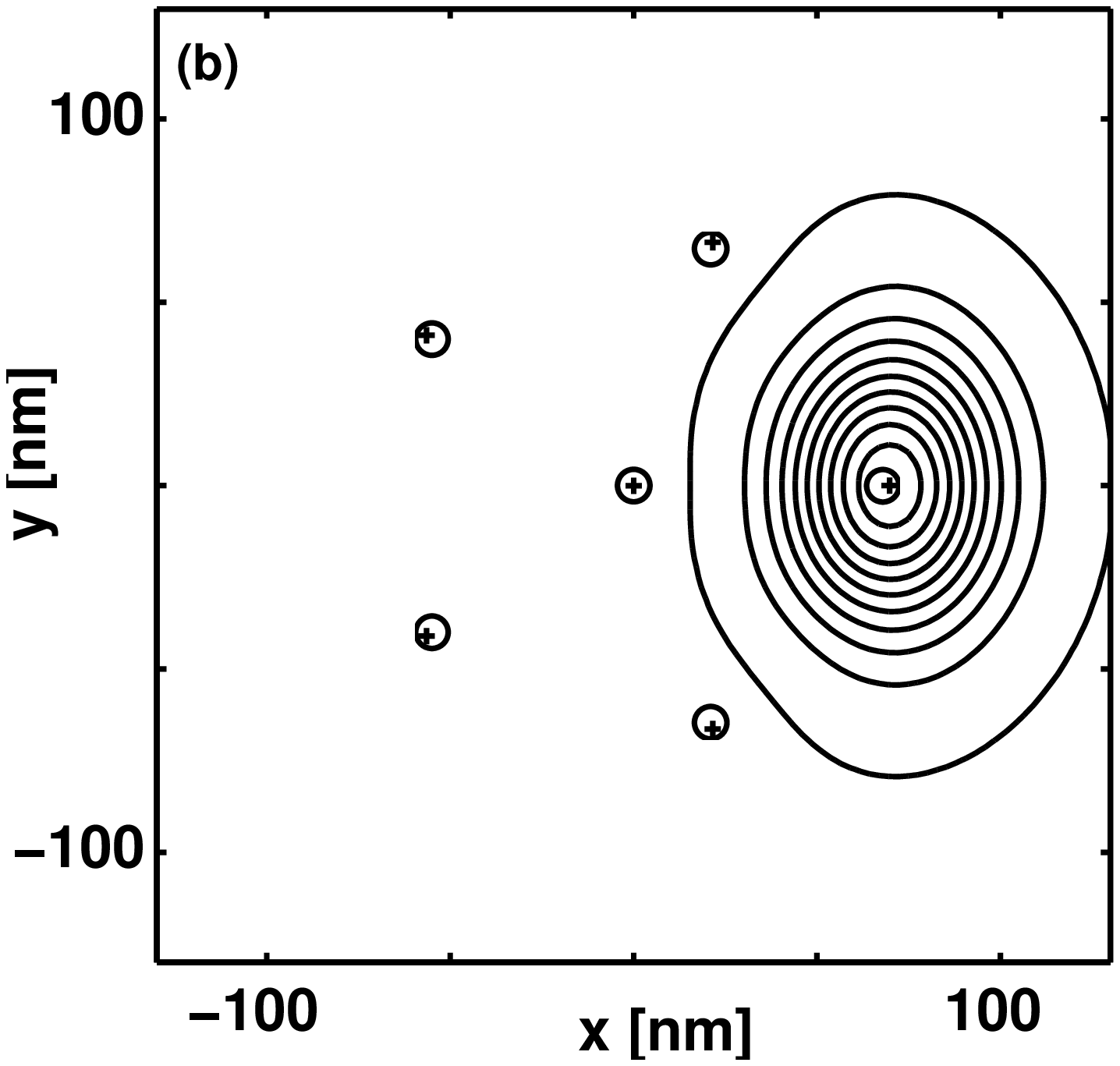}\\
  \includegraphics[width=0.425\columnwidth]{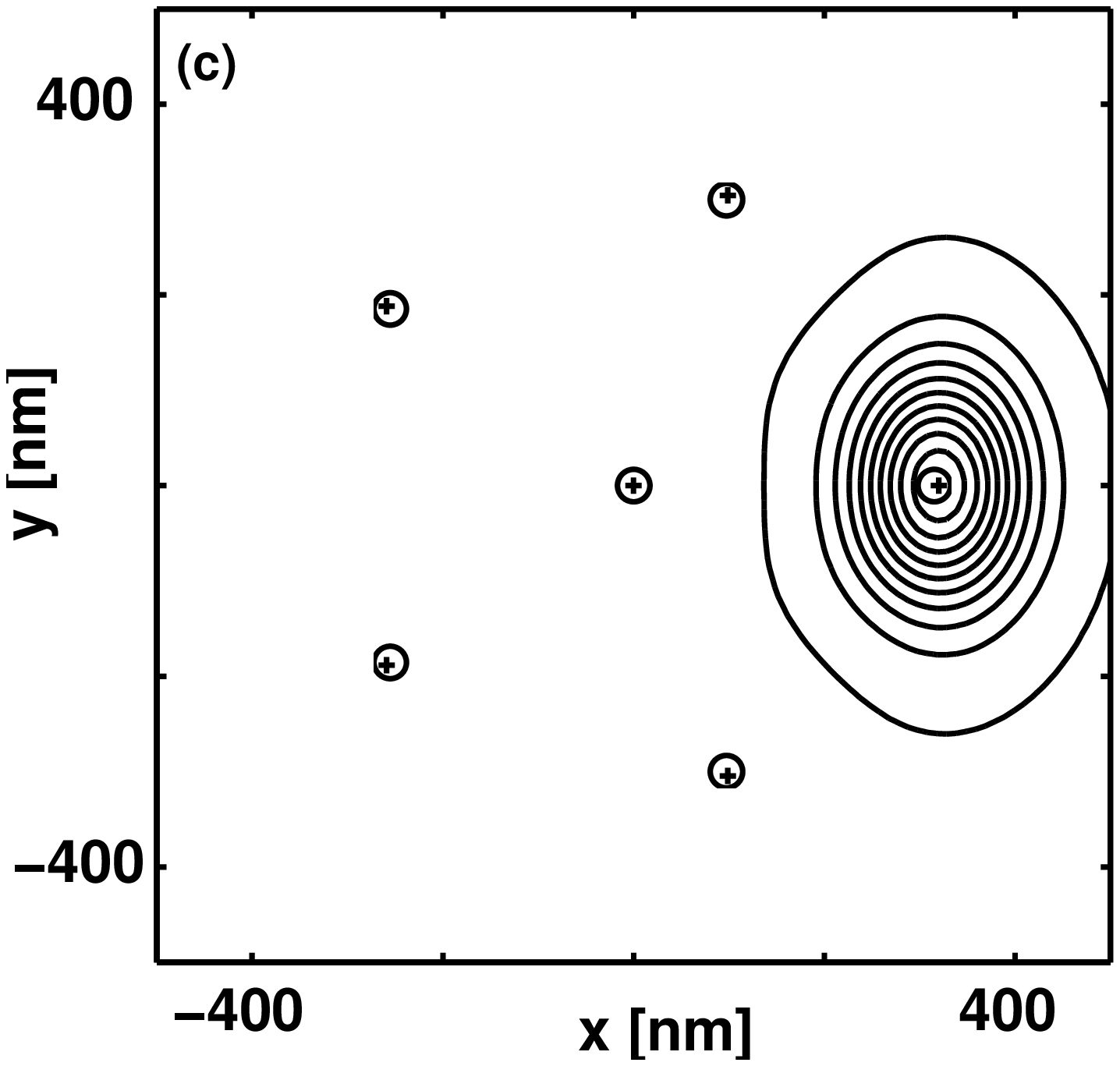}
  \includegraphics[width=0.425\columnwidth]{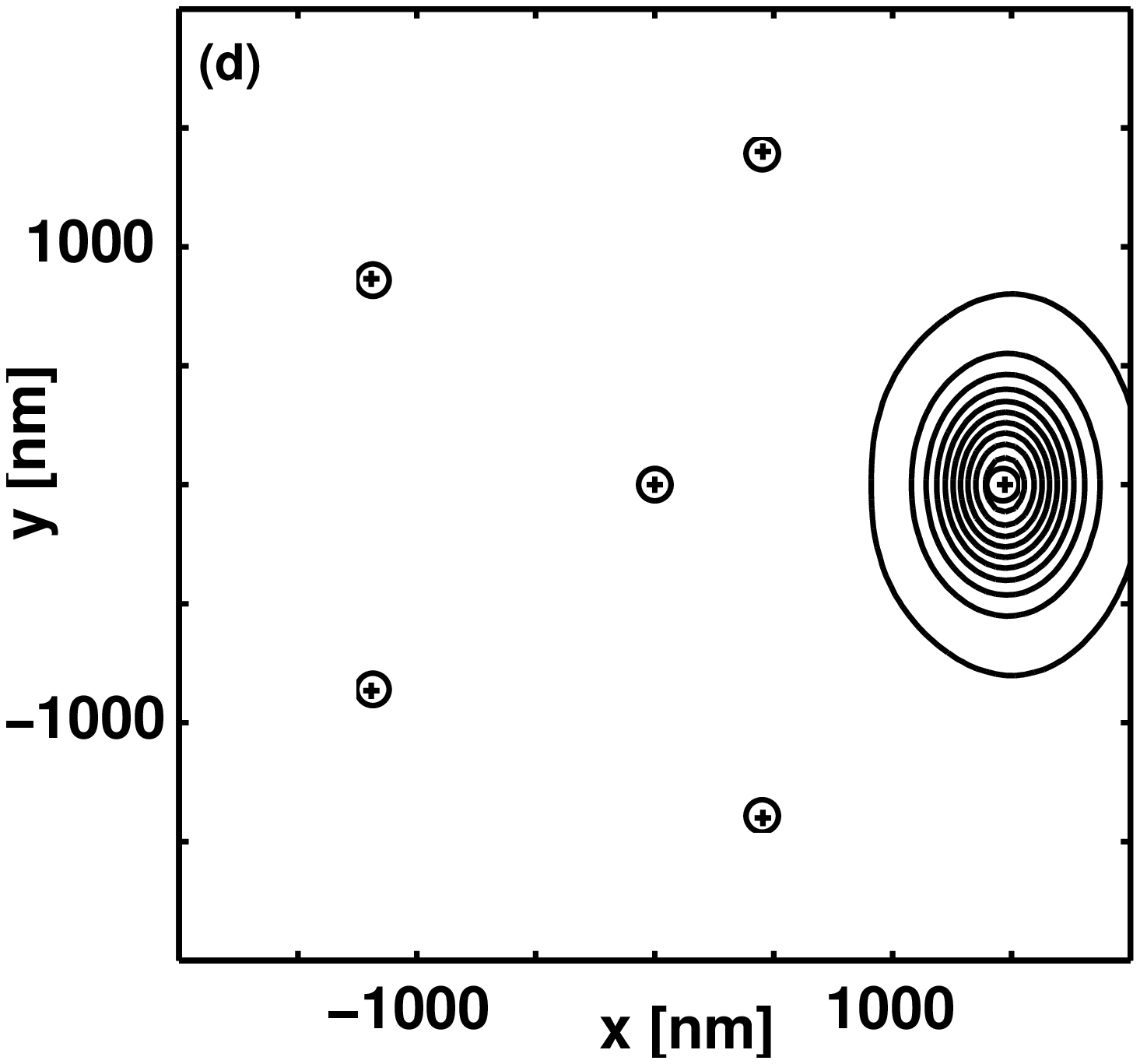}
\end{center}
\caption{Conditional probability density for the right-most
  electron. The contours are spaced uniformly from 0.01 to 0.91.  We
  mark with a plus the most probable electron positions, and with a
  circle the classical positions. The confinement strength $\hbar
  \omega$ is: (a) $10$~meV, (b) $1$~meV, (c) $0.1$~meV, and (d)
  $0.01$~meV.}
\label{WC}
\end{figure}

The most probable configuration ${\bf R}^*$, found by maximizing the
density $| \Psi({\bf R}) |^2$, should approach in the limit of weak
confinement the classical electron positions. This is not, however,
enough to show that the system is close to a classical one. One can
study the quantum fluctuations very conveniently using the conditional
single-particle probability distribution $\tilde \rho ({\bf r})$,
defined as
\begin{equation}
\tilde \rho ({\bf r}) = \left | \frac{\Psi({\bf r},{\bf
r}^*_2,\dots,{\bf r}^*_{N})}{\Psi({\bf r}_{1}^{*},{\bf
r}^*_2,\dots,{\bf r}^*_{N})} \right |^2 \ ,
\label{rhod}
\end{equation}
where the coordinates ${\bf r}_i^*$ are fixed to the ones from the
most probable configuration ${\bf R}^*$. In the classical limit, the
density $\tilde \rho ({\bf r})$ is more and more peaked around the
classical position ${\bf r}_1^*$, but still shows quantum
fluctuations.  This is shown in Fig.~\ref{WC}

The calculation of $\tilde \rho$ is very easy in VMC.  One should
first, of course, find the most probable electron positions.  In doing
this, the gradient of the wave function (also needed for the
calculation of the local energy, and for this reason usually done
analytically) is very useful. After that, one moves the ``probe
electron'' to all points where the value of $\tilde \rho$ is wanted,
and evaluates the ratio of wave functions as in Eq.~(\ref{rhod}).
This ratio is automatically done while sampling the configurations in a
VMC simulation.  One should also notice that $\tilde \rho$ does not
contain noise unlike many more common VMC observables, such as the
density or the radial pair distribution function.

\begin{figure}[hbt]
\begin{center}
  \includegraphics[width=0.75\columnwidth]{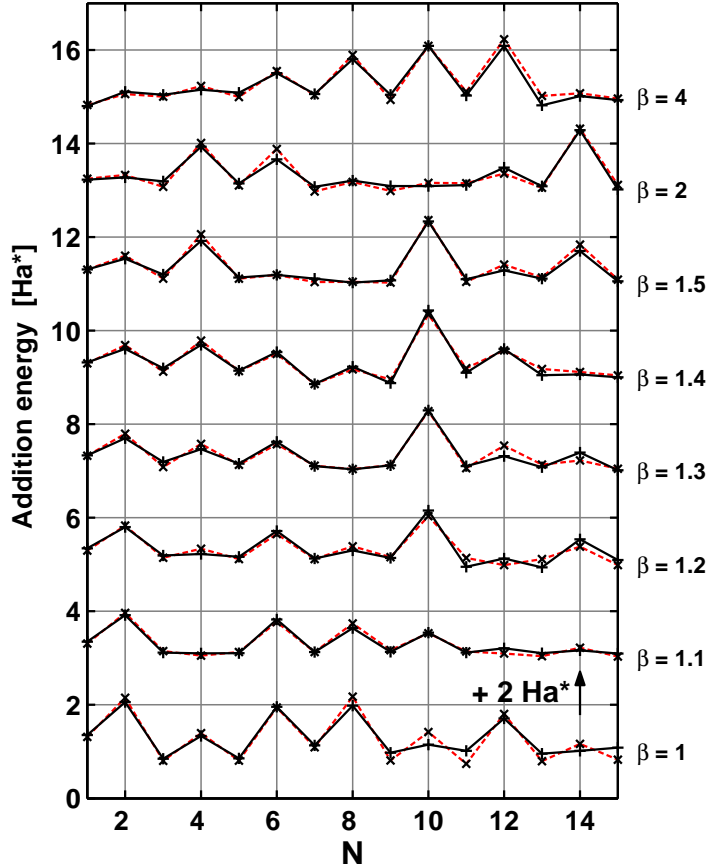}
\end{center}
\caption{Addition energy spectra for rectangular quantum dots with
  different axis ratio $\beta$.  The density-functional theory and VMC
  results are given by pluses and crosses connected with solid and
  dashed lines, respectively.}
\label{EVa}
\end{figure}

\begin{figure}[hbt]
\begin{center}
  \includegraphics[width=0.99\columnwidth]{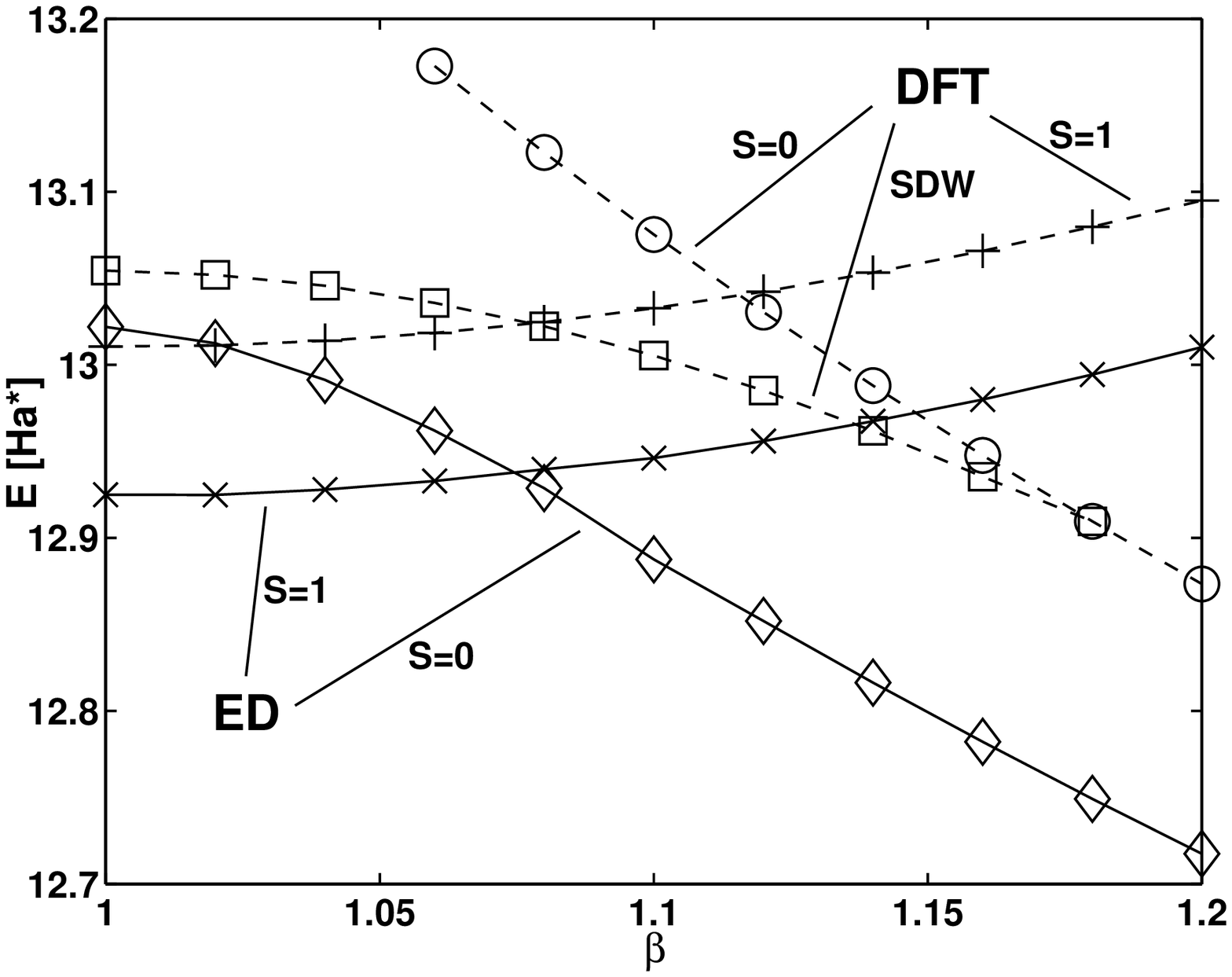}
\end{center}
\caption{Energy of the four-electron dot as a function of $\beta$. The
  solid lines present exact energies, we use crosses for $S=1$ and
  diamonds for $S=0$, correspondingly. The dashed lines are
  density-functional theory energies, pluses for $S=1$, boxes for the
  $S=0$ broken-symmetry solution, and circles for the symmetric $S=0$
  energy.}
\label{EVb}
\end{figure}

It is rather surprising that non-interacting single-particle states
can be used for rectangular hard-wall QD's also\cite{recta}. Our VMC
results using these agree very well with the density-functional theory
data\cite{recta,AetC}. To see this, the addition energies $\Delta(N)$
(defined as a second difference in total energies
$E[N+1]-2E[N]+E[N-1]$) are plotted in Fig.~\ref{EVa}.
The differences between the results of the two different methods are
extremely small in all cases. The data shows that by tuning the ratio
of the two axis of the system, the shell structure of the electronic
structure can be tuned in a controllable fashion. This is one of the
aims in novel microelectronics, namely that one can engineer the
properties of the nanoscale semiconductor system to suit the need at
hand. This parameter-tuning is, of course, not possible in a simple
way in the cases of atoms and molecules. The parameter-tuning also has
an interesting effect on the theoretical calculations. One can study
how various approximations perform as a function of the change in the
system parameter. This is done in Fig.~\ref{EVb},
 where we compare
the total energies from an exact diagonalization to a
density-functional one for the case of a four-electron rectangular
QD. The parameter we change is the side-length ratio $\beta$. The
mean-field approach has a broken-symmetry solution, and our analysis
shows that the energy lowering due to breaking the symmetry is too
large, and in addition, the broken-symmetry solution results from the
fact that the system naturally consists of two important
configurations\cite{AetC}.  It might be possible to treat several
configurations by a mean-field approach, see Ref.~\onlinecite{AetC}
for a suggestion, but it is not trivial.

\begin{figure}[hbt] 
\begin{center}
 \includegraphics[width=0.99\columnwidth]{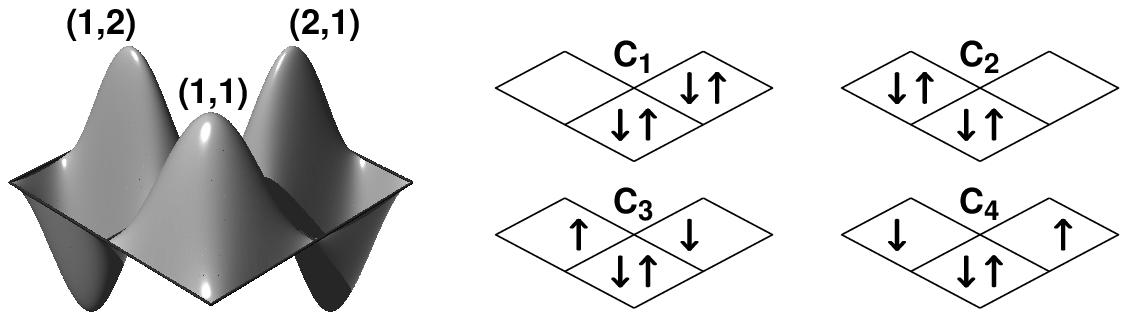}
\caption{Left panel: The three lowest single-particle states and their
quantum numbers ($n_x$,$n_y$). The (1,1) state is a positive sine
function and both of the two higher states have one node (on $x$ and
$y$ axis, correspondingly). Right: Electron occupations for the four
important $S_z=0$ configurations $C_i$.}
\label{foofoo}
\end{center}
\end{figure}

One should note that also in VMC, it is very difficult to know
beforehand if one needs a multi-configuration wave function or
not. Some analysis can be performed based on the single-particle
energetics. However, even in cases where one can form several
different configurations that are close in energy, it is not clear
that all these configurations are relevant for the ground state. Of
course, symmetry plays again role here. As a concrete example, one can
compare the four-electron $S=0$ states of a parabolic and a square
hard-wall QD. The single-particle states of the parabolic case can be
obtained from Eq.~(\ref{simplewf}), and the ones for the square one in
Fig.~\ref{foofoo} with the relevant ground-state configurations.
The relevant configurations for the parabolic case can be found in a
similar fashion from the three lowest single-particle states (by
replacing the ($n_x$,$n_y$) states by the lowest ($n$,$l$) ones). From
the four configurations, one can form both $S=1$ and $S=0$ states with
$S_z=0$. The $S=1$ state with $S_z=0$ consists of the configurations
$C_3$ and $C_4$ with equal weight.  This is degenerate with the
$S=S_z=1$ state, which is a single determinant. This configuration can
be obtained from $C_3$ or $C_4$ by flipping the down-spin from the
higher state. One can, however, find a truly two-configuration
$S=S_z=0$ state from the configurations $C_1$ and $C_2$. Without the
electron-electron interaction, the states $\Psi_{\pm}=C_1 \pm C_2$
have the same energy. However, when the interaction is present, the
degeneracy of these states is split, and one of them is lowered in
energy. This can directly be seen from the Hamiltonian matrix on the
basis of these two configurations
\begin{equation}
H=\left(
\begin{array}{cc}
E_1 & \delta  \\
\delta & E_2 
\end{array} \right) \ ,
\end{equation}
where the energies $E_i$ are for configurations $C_i$ and $\delta$
denotes the coupling of the configurations. If one performs the same
analysis for the two configurations of the parabolic case, one
actually finds that the coupling between the two configurations is
zero ($\delta=0$).  This means that the configurations do not mix. The
reason for this is symmetry. One can find this out by calculating the
Coulomb matrix element between the configurations, but it is also
possible to see this directly from symmetry. The two configurations in
the parabolic case correspond to ground states with angular momentum
$\pm 2$. As the problem has circular symmetry, the angular momentum is
a good quantum number both for the single- and many-particle
states. In this way the configurations with different symmetry do not
mix. One should note that here we have used real wave functions for
the square QD and complex for the parabolic one. If one would use real
ones for the parabolic case, several configurations are
needed\cite{cyrus}. However, the use of complex wave functions in a
square case does not bring the wave function to a single-configuration
one.

\subsection{Integral Quantum Hall Effect Regime}\label{iqhe}

The finite $B$ lowers the symmetry of the parabolic QD, as the states
that form shells at $B=0$ are not degenerate anymore. This can be seen
in Fig.~\ref{QD_states_5meV}. The reason for this is the ``rotation''
term in the Hamiltonian, discussed in Section~\ref{junk}, that lowers
the energy of states that rotate to the correct direction, and highers
the ones with an angular momentum of opposite sign.

\begin{figure}[hbt] 
\begin{center}
 \includegraphics[width=0.7\columnwidth]{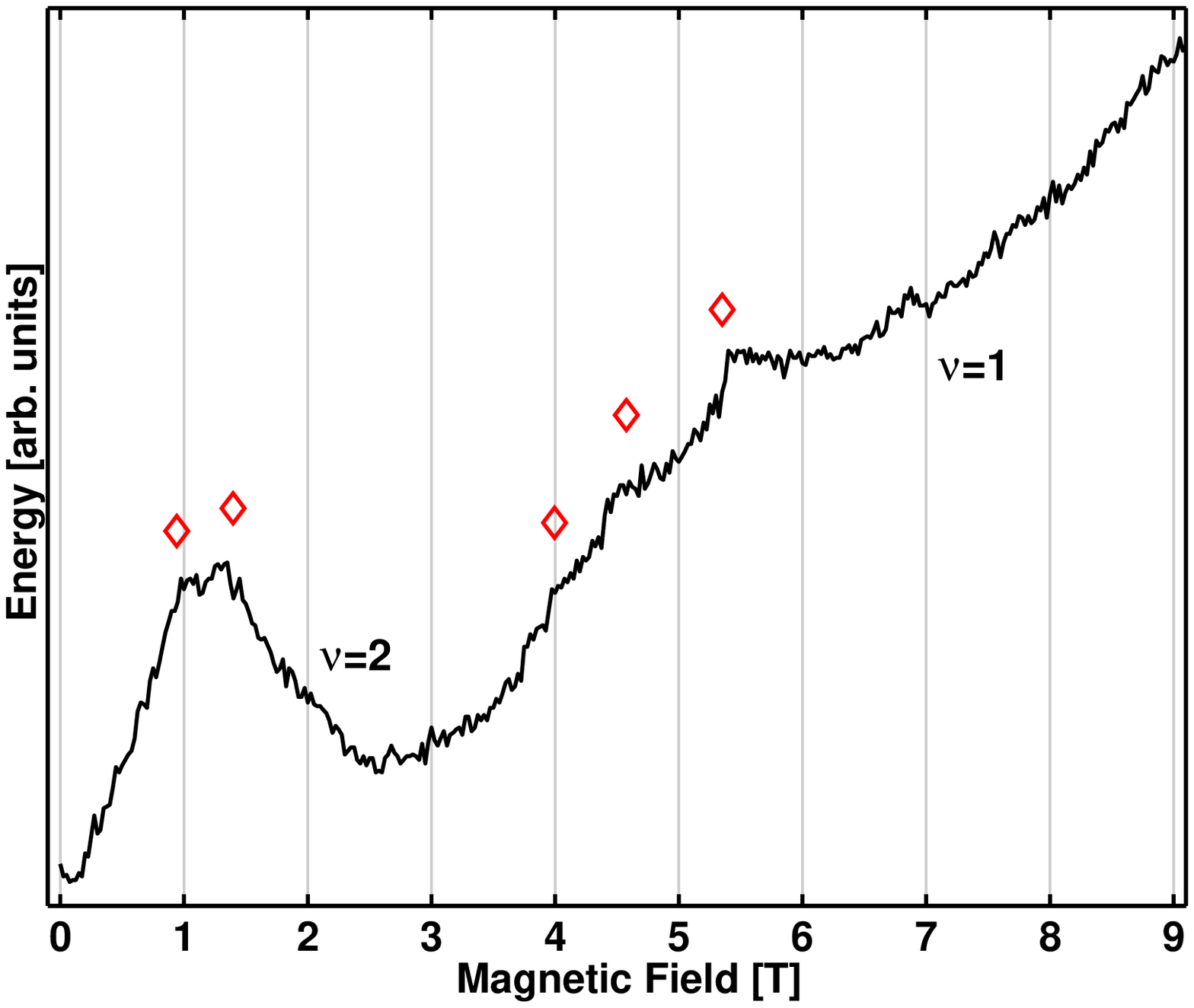}\\
 \includegraphics[width=0.9\columnwidth]{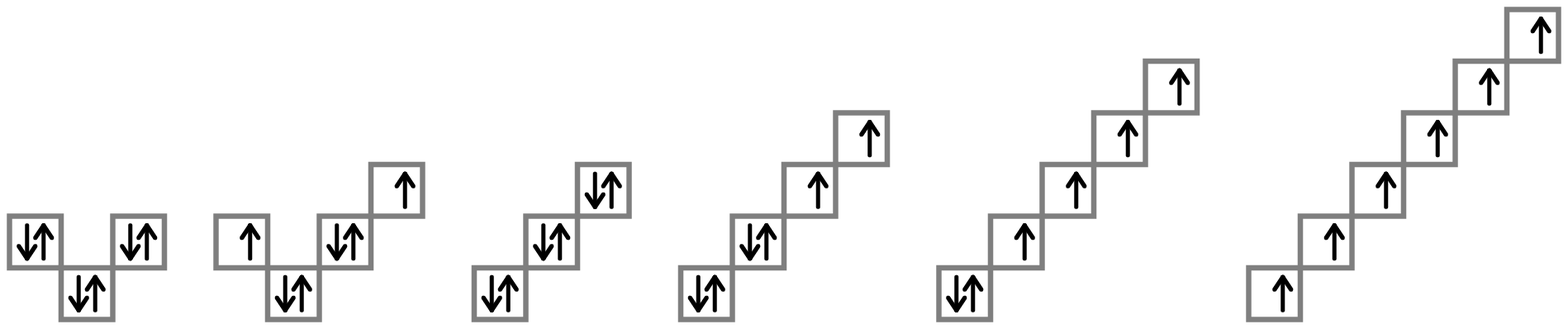}
\caption{{Upper panel:} Experimental energy as a function of the
  magnetic field $B$ for $N=6$ QD. The kinks show the transition
  points where the lowest energy state changes.  The $B$-values of the
  VMC transition points are marked with '$\diamond$' (vertical
  positions arbitrary). The two ground-state regions that are related
  to quantum Hall states with filling factors $\nu=1$ and 2 are
  shown. {Lower panel:} Electron occupations of the lowest energy
  states.}
\label{Tjerk}
\end{center}
\end{figure}
A single-determinant-Jastrow wave function is still accurate in the
IQHE regime\cite{weakPRB}. One can also find good agreement with the
experimental transition points\cite{weakPRB}. This can be seen in
Fig.~\ref{Tjerk}, where the curve shows the measured data, with kinks
at the transition points, and the markers shows the $B$-values where
the VMC finds a change in the ground state configuration.  The VMC
parameters used are $m^*/m_0=0.067$, $\epsilon=12.9$,
$\hbar\omega_0=4.5$~meV, g$^*=-$0.44, and
$V_{ee}=\alpha\frac{e^{2}}{\epsilon r_{ij} }$ with $\alpha=0.7$.
Basically, we have two parameters to tune, but we have tried to
extract them from the experimental data.  The confinement value used
is a reasonable one compared with the experimental value of
$\hbar\omega_0\approx 5$~meV for the one-electron dot, as the
confinement is weaker for larger electron numbers. The scaling of the
Coulomb interaction value $\alpha=0.7$ is obtained from the
experimental transition point for the $N=2$ case, assuming that the
confinement strength is nearly the same as for one-electron QD. The
choice of $\alpha$ means that the electron-electron interaction is
only 70 percent of what one would assume from the material
parameters. There are two effects that cause this. The first is the
softening of the interaction at small electron-electron
distances. This is due to the fact that in our model we completely
neglect the third spatial dimension. In a realistic case, when the
electrons approach each other, they can use this dimensionality to
avoid each other. This results in an interaction potential that is
effectively softer at small electron-electron distances. One
possibility to insert this in to theory would be to use an interaction
of the type
\begin{equation}
V(r_{ij})=\frac{C}{\sqrt{\delta^2+r_{ij}^2}} \ ,
\end{equation}
where $\delta$ is directly proportional to the finite width of the
system.  The second effect is that in experimental setups, the
surrounding metallic gates can screen the electron-electron
interaction at large inter-electron distances. One can use different
approximations for the screening, but qualitatively, the effect of
this is also that the interaction becomes weaker.  We have, in some
approximation, taken these into account by tuning the interaction by
the parameter $\alpha$. What makes this approach more favorable than
the ones that would take into account the two experimental facts
discussed above is that now we have only one parameter to tune,
instead of one for the finite thickness and one for the screening.

The lower part of Fig.~\ref{Tjerk} shows the single-particle
occupations in the determinant. The right-most one corresponds to MDD,
and there all spins are up and occupy LLL states with
$l=0,1,\dots,5$. In the left-most configuration, electrons fill the
two lowest shell of $B=0$. The logic of the configurations between
these two extremes is that the occupation is moving towards LLL. The
second configuration is obtained from the first one by flipping one
spin and moving it to LLL. The reason behind having $S=1$ is the
exchange energy, and only secondly the energy lowering by the Zeeman
term. In the third configuration, the electrons are already on the
LLL, and this state corresponds to the $\nu=2$ IQHE state. As $B$
increases, the electrons start to flip spins one by one, until the
configuration is the fully spin-polarized $\nu=1$ state. One can see
that the determinant part of the wave function is here given by
Eq.~(\ref{lllwf}). These determinants have a CM motion in the ground
state. In some cases that have more than one important non-interacting
configuration present in the interacting case, one can still transform
it to a single-configuration one by a coordinate transformation
$\mathbf{r} \rightarrow
\mathbf{r}-\mathbf{r}_{\mathrm{cm}}$\cite{weakPRB}.  This is done only
on the phase part of the wave function, as the exponential part is
always the same and corresponds to the correct ground state of the CM
motion. This coordinate transformation has an important consequence,
namely that after this is done, every single-particle wave function in
the determinant depends on all electron coordinates via
$\mathbf{r}_{\mathrm{cm}}$. In practice, this means that the commonly
used computational rules for the determinant ratios are not trivially
applicable anymore. For this reason, it might be better to move all
the electrons at the same time.

It is relatively easy to find the relevant single-particle
configurations, like the ones shown above in Fig.~\ref{Tjerk}. For a
weak $B$, the configurations reflect the shell structure of the $B=0$
case where Hund's rules can be used as an initial guideline. When one
moves to higher $B$ values, the total angular momentum gets
larger. The energies plotted in Fig.~\ref{QD_states_5meV} give some
idea of various possibilities for the configurations. One should also
note that it is rather trivial to test various configurations in VMC,
as the simulations are very fast. In addition, even the energies with
rather large statistical error are accurate enough to tell if one
should pay any attention to the configuration in question.

One should note that in this regime, also the mean-field theories
might work in a reasonable fashion in most cases. The
density-functional theory predictions are compared with the those of
VMC in Refs.~\onlinecite{epjb} and \onlinecite{testDFT}. The cases
where mean-field theory breaks down are typically such that there are
several important configurations in the ground state. Another minor
problem of the mean-field results is that the relative accuracy
depends on the total spin of the system\cite{testDFT}.

\subsection{Fractional Quantum Hall Effect Regime}

The FQHE regime starts at the reconstruction of the MDD state. The
important difference, as pointed out above, is that in the FQHE
regime, the interactions start to dominate. At IQHE cases the phase
structure is still very well approximated by a single Slater
determinant, but for the FQHE limit, the many-body wave functions have
phase structures that are fundamentally of a many-body nature.  The
rather accurate Laughlin states are among the few
exceptions\cite{Laughlin_13}. These wave functions for QD's can be
written as
\begin{equation}
\Psi(\mathbf{r}_1,\mathbf{r}_2,\dots,\mathbf{r}_N) =
\exp\left[-\sum_{i=1}^N r_i^2/2\right] \prod_{i<j}^N z_{ij}^m \ ,
\end{equation}
with only one letter ($m$, needs to be odd) difference to the MDD
state, see Eq.~(\ref{mddwf}). Actually, MDD is obtained with $m=1$ and
those with $m=3,5,\dots$ really correspond to a FQHE case. The FQHE
filling factor in these states is $\nu=1/m$. If one splits the
MDD-part from the Laughlin states, the remaining factor $\prod_{i<j}^N
z_{ij}^{m-1} $ can be thought as a generalized two-body Jastrow
factor.

One can multiply the Laughlin wave function by a traditional Jastrow
factor, and lower the energy. One can thus split the correlation
effects into two parts, the one on the LLL and the one that takes into
account the higher Landau levels. We call the second part of
the correlation effects Landau-level mixing (LLM).  For three
electrons, a particle number that is still rather accurately solved by
the exact diagonalization, VMC results show that 96-98\% of the LLM
effects on the energy are captured by the simple two-body Jastrow
factor\cite{epl}.

It turns out that the LLL part in the correlation is far more
complicated. The two-electron case is, of course, a simple exception,
as the LLL wave functions are easily found from Eq.~(\ref{2e}).  But
for larger particle numbers, no simple routes to exact wave functions
have been found.  The LLL wave function of fully spin polarized cases
in the FQHE regime can be written as
\begin{equation}
\Psi(\mathbf{r}_1,\mathbf{r}_2,\dots,\mathbf{r}_N) =
\exp\left[-\sum_{i=1}^N r_i^2/2\right] \prod_{i<j}^N z_{ij} \, P_{\Delta L} \ ,
\label{PsiP}
\end{equation}
where $P_{\Delta L}$ is a symmetric polynomial of degree $\Delta L$ of
all the variables $\{z_i\}_{i=1}^N$.  The wave function is thus the
MDD one multiplied by a polynomial.  The task is to find the
coefficients of this polynomial.  If $\Delta L=1$, the polynomial is a
trivial sum
\begin{equation}
S_1=\sum_{i=1}^N z_i \ ,
\end{equation}
which can easily be seen to correspond to an excitation of the CM
motion, as $S_1=N z_{\mathrm{cm}}$. The case of $\Delta L = 2$ is also
easy to find, if one writes it as a polynomial that is orthogonal to
$S_1^2$. To do this in practice, one can write any $P_2$ and apply a
coordinate transformation (as we did in Section \ref{iqhe}) $z_i
\rightarrow z_i- z_{\mathrm{cm}}$. If the result is zero, then the
initial polynomial was $S_1^2$. In other cases one finds a polynomial
which is
\begin{equation}
\tilde S_2 = \sum_{i_1<i_2}^N (z_{i_1}-z_{\mathrm{cm}})
(z_{i_2}-z_{\mathrm{cm}}) \ ,
\end{equation}
where the tilde on $\tilde S_2$ is used to distinguish it from the
elementary symmetric polynomials
\begin{eqnarray}
S_1 &=& \sum_{i_1} z_{i_1} \ , \nonumber \\
S_2 &=& \sum_{i_1<i_2} z_{i_1} z_{i_2} \ , \nonumber \\
S_k &=& \sum_{i_1<i_2<\dots<i_k} z_{i_1} z_{i_2} \dots
         z_{i_k} \ ,\\ \label{lllbasis}
S_N &=& z_{i_1} z_{i_2} \dots z_{i_N} \ . \nonumber
\end{eqnarray}
These functions can actually be used as a basis for the QD studies in
the FQHE regime\cite{StabilityMDD,MCD}. If one is only interested in
the ground states, the $S_1$ polynomial can be dropped from the basis,
but then one has to replace $z_i$ by $z_i- z_{\mathrm{cm}}$ in the
higher order polynomials. We mark these polynomials by $\tilde S_L$
(with $L>1$) below.

Using the polynomials $\tilde S_L$, the problem for three electrons is
rather low dimensional. The calculations show that the ground states
occur at additional angular momentum values $\Delta L = 3 \times i$, where
$i$ is a positive integer\cite{epl}. The LLL-part of the ground state
with $\Delta L=3$ is directly $\tilde S_3$, as it is the only
possibility on the given basis. One has to remember that $S_1$ cannot
be present in ground states, as it corresponds to a CM excitation. For
$\Delta L = 6$, the two polynomials in the basis are $\tilde S_3^2$
and $\tilde S_2^3$.  The polynomial $\tilde S_3^2$ is more important,
it has an overlap of 0.958 with the exact state, whereas the Laughlin
$m=3$ state has an overlap of 0.991. The next ground state with
$\Delta L = 9$ consists of $\tilde S_2^3 \tilde S_3$ and $\tilde S_3^3
$, where the second one is more important with an overlap of 0.972
with the exact state. A similar trend continues for higher values of
$\Delta L$, and still at $\Delta L =30$ when the full basis (using $S$
instead of $\tilde S$) is 91 dimensional, the problem is six
dimensional in the basis of $\tilde S_L$ functions. Furthermore, the
overlap of $\tilde S_3^{10}$ with the exact state is still 0.983.

\begin{figure}[hbt]
\begin{center}
\includegraphics[width=0.79\columnwidth]{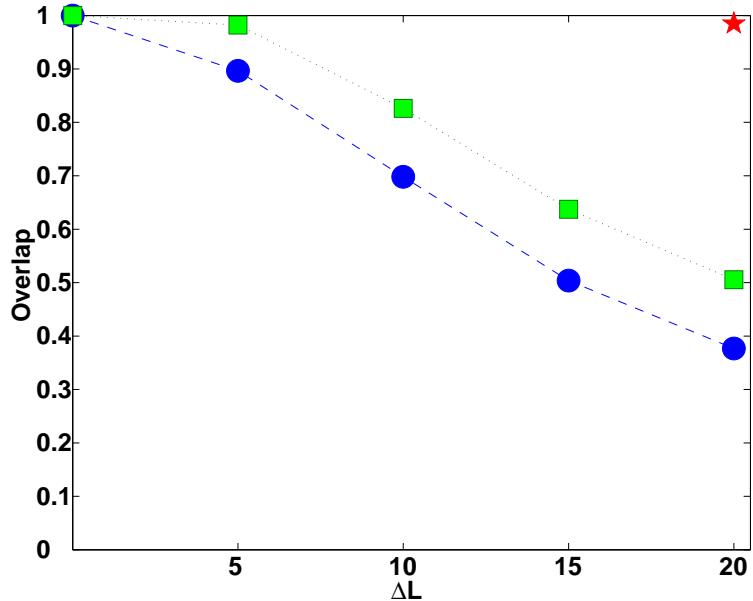}
\end{center}
\caption{Overlap of a single-determinantal wave function with the
exact one as a function of the additional angular momentum $\Delta L$
for five electrons.  The higher points are obtained with a coordinate
transformation $z_i \to z_i- z_{\mathrm{cm}}$, and the lower ones
without it. The pentagram at $\Delta L = 20 $ shows the overlap of the
Laughlin $\nu=1/3$ state.}
\label{overlaps}
\end{figure}
\begin{figure}[hbt]
\begin{center}
  \includegraphics[width=0.99\columnwidth]{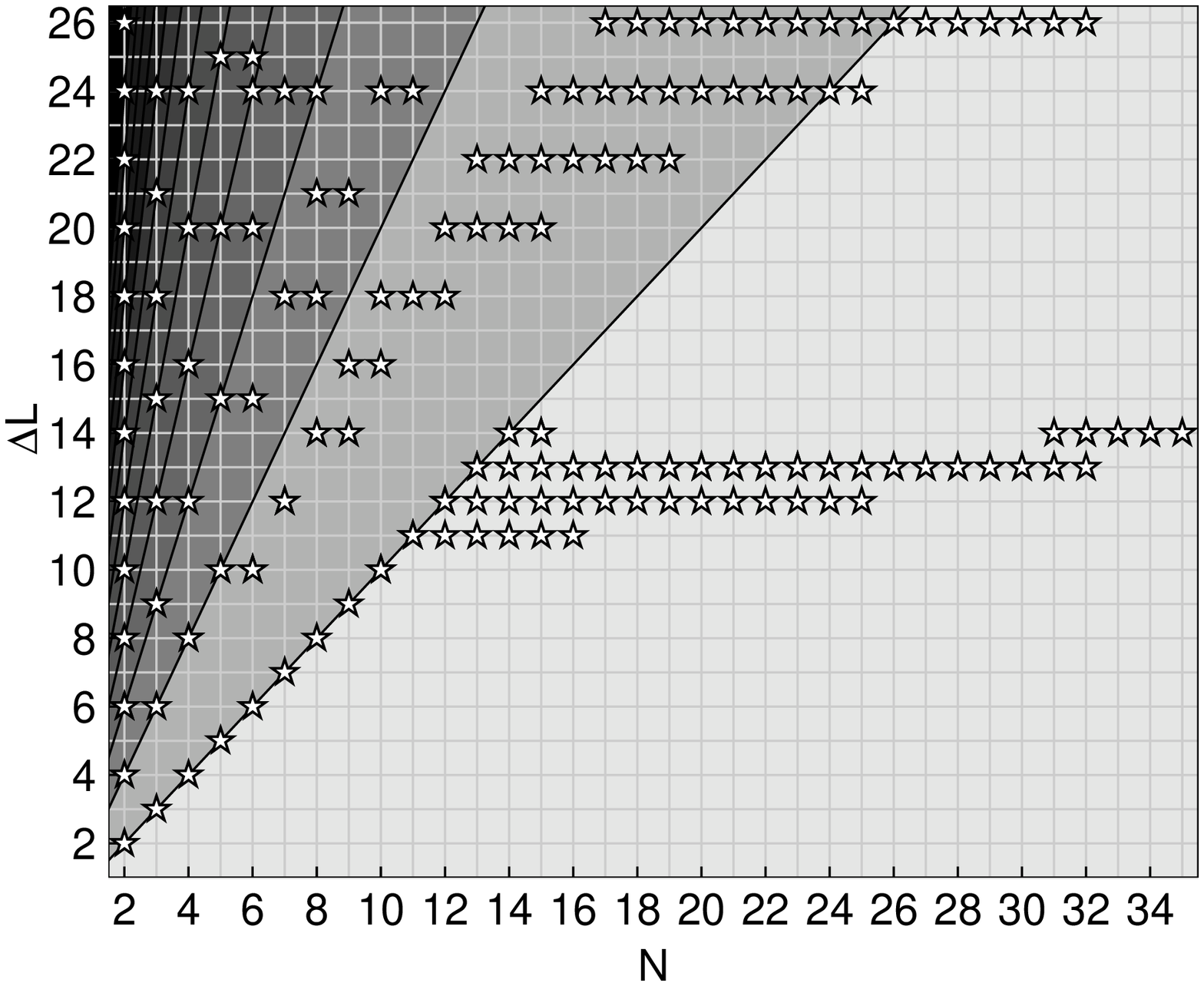}
\end{center}
\caption{Possible ground state additional angular momentum values
($\Delta L$) for various particle numbers ($N$) from an exact
lowest-Landau-level theory. For small $N$, $\Delta L$ is regular but
more complicated structures are seen on larger particle numbers.}
\label{GSs}
\end{figure}

For larger particle numbers, the physics get more complicated. One can
approximate the determinant part by a single determinant written in
coordinates $z_i- z_{\mathrm{cm}}$. In Fig.~\ref{overlaps}, the
overlap for the $N=5$ case is shown. One can see that the overlap
decreases dramatically for large values of $\Delta L$.  For $\Delta L
= 5$, one can approximate the state by a polynomial $\tilde S_5$. For
higher $\Delta L$, more and more polynomials are needed to accurately
describe the states.  In spite of that, one can actually find very
nice agreement for the $\Delta L$ values of the ground states using
transformed determinants\cite{beyondMDD}. One would not expect this
from the overlaps shown above.  The values of $\Delta L$ are not
actually trivial for larger particle numbers. If one forgets the LLM
and finds the possible ground states on LLL theory, exact
diagonalization is a very natural tool for that\cite{vortex}. The LLL
ground states can be seen in Fig.~\ref{GSs},
where we plot by a star each $\Delta L$ value that can be the
ground state of the system of $N$ electrons at some value of $B$.  For
$N\le5$, $\Delta L$ is an integer times $N$, but for larger particles
number, more irregular behavior of $\Delta L$ is seen. The state with
$\Delta L = N$ is a possible ground state for $N \le 14$. For $N=14$,
the ground state has an overlap of 0.996 with a state with polynomial
$\tilde S_{14}$. One should note that this polynomial corresponds to a
case with one flux quantum through the center of mass.  The state of
Ref.~\onlinecite{diagLLL} is nearly exact with an overlap of 0.99991
with the exact state. The nature of the first post-MDD state changes
at $N=12$. For $N\ge 15$ the $\Delta L = N$ is not a ground state
anymore.

The composite-fermion (CF) theory is one possibility for constructing
the variational wave function\cite{jain97,SamiDI}. It is, however,
rather nontrivial to construct a complete CF theory for all states of
the QD's in the FQHE regime.  Especially, one needs to go beyond the
mean-field rule used in many cases to obtain the $\Delta L$ values of
the CF theory\cite{beyondMDD}.  In an easy example of the CF theory,
two vortices are added to each electron as
\begin{equation}
\prod_{i<j}^N z_{ij}^2 \ ,
\end{equation}
and the electrons are considered to move in an effective magnetic
field that is obtained from the external $B$ by subtracting the field
strength of the added vortices. If the resulting field is still
reasonably strong and to the direction of the external $B$, the
electron wave function can be taken to be, e.g, that of the MDD. Doing
this, the $\nu=1/3$ Laughlin state is obtained. If, however, the
resulting effective field is again reasonably strong but this time
opposite to the external $B$, the electrons can form the MDD for this
opposite $B$. To obtain this, one has to transform $z_{ij} \to
z_{ij}^*$ in MDD wave function. Next, one needs to project this to
LLL, and one possibility is to replace $z^*$ by a derivative $\partial
/ \partial z$. Thus the wave function is
\begin{equation}
\Psi(\mathbf{r}_1,\mathbf{r}_2,\dots,\mathbf{r}_N) =
\exp\left[-\sum_{i=1}^N r_i^2/2\right] 
\prod_{i<j}^N 
\left(
\frac{\partial}{\partial z_{i}}-
\frac{\partial}{\partial z_{j}} 
\right)
\prod_{i<j}^N z_{ij}^2 \ ,
\end{equation}
and the two products can be simplified to a simple product of MDD:
$\prod z_{ij}$. A trivial way to find this is to note that MDD is
the only LLL state with the given angular momentum, and that each
derivative lowers the angular momentum of $\prod z_{ij}^2$ by one.
These two examples are the limiting ones, and there are other
possibilities in between\cite{SamiDI}. The wave functions of the
electrons moving in effective fields can be as complicated as the ones
of the original problem. In addition, the projection used above is not
the only possibility. An interesting finding using CF theory is that
one can obtain more accurate states than the Laughlin $\nu=1/3$ one by
calculating
\begin{equation}
\Psi(\mathbf{r}_1,\mathbf{r}_2,\dots,\mathbf{r}_N) =
\exp\left[-\sum_{i=1}^N r_i^2/2\right] 
\prod_{i<j}^N 
\left(
\frac{\partial}{\partial z_{i}}-
\frac{\partial}{\partial z_{j}} 
\right)
\prod_{i<j}^N z_{ij}^4 \ .
\end{equation}
This means that it is energetically favorable to add four vortices to
each electron and move on a field that is opposite to the original one
than to add two vortices and have an effective field that is pointing
to same direction than the original field.
In addition to Ref.~\onlinecite{jain97}, one can find see more
details in Ref.~\onlinecite{SamiDI}.

An interesting question is how the LLM affects the LLL part of the
wave function. In other words, if one finds first the polynomial of
Eq.~(\ref{PsiP}) on LLL theory, the important question is how much do
the optimal coefficients change when one inserts the Jastrow
correlation factor in the wave function. A possible method to answer
this is the Monte Carlo -based diagonalization (MCD)\cite{MCD}. The
method is similar to the exact diagonalization, but the expansion is
done on a correlated basis. This means that the expansion converges
faster than the one done on the non-interacting configurations. One
can find rather simple form for the Hamiltonian matrix elements,
namely
\begin{equation}
h_{ij}=\left \langle P_i^*P_j \times 
\left\{ \frac 12 (E_i + E_j) + V_I 
+ \frac 12 \left|\frac{\nabla J}{J}\right|^2
\right\}
\right \rangle_{|\Psi_0 J|^2} \ ,
\end{equation}
where configurations $\mathbf{R}$ are sampled from $|\Psi_0 J|^2$, and
in our case, $\Psi_0$ is MDD, $P$'s are the symmetric polynomials,
$V_I$ is the interaction and impurity potential, and energies $E_i$
and $E_j$ are found from
\begin{equation}
\mathcal{H}_0 \Psi_0 P_i =  E_i \Psi_0 P_i \ .
\end{equation}
The overlap matrix elements are 
\begin{equation}
s_{ij}=\left \langle P_i^*P_j
\right \rangle_{|\Psi_0 J|^2} \ ,
\end{equation}
and the Schr\"odinger equation is given in a matrix form as
\begin{equation}
\mathbf{H}\mbox{\boldmath $\alpha$}=E \mathbf{S} \mbox{\boldmath
  $\alpha$} \ .
\end{equation}
The reason to use Monte Carlo for the matrix elements is that the
integrals needed are high-dimensional. By the MCD method, one obtains,
in addition to the ground states, also the excited states.

\begin{figure}[hbt]
  \includegraphics[width=0.49\columnwidth]{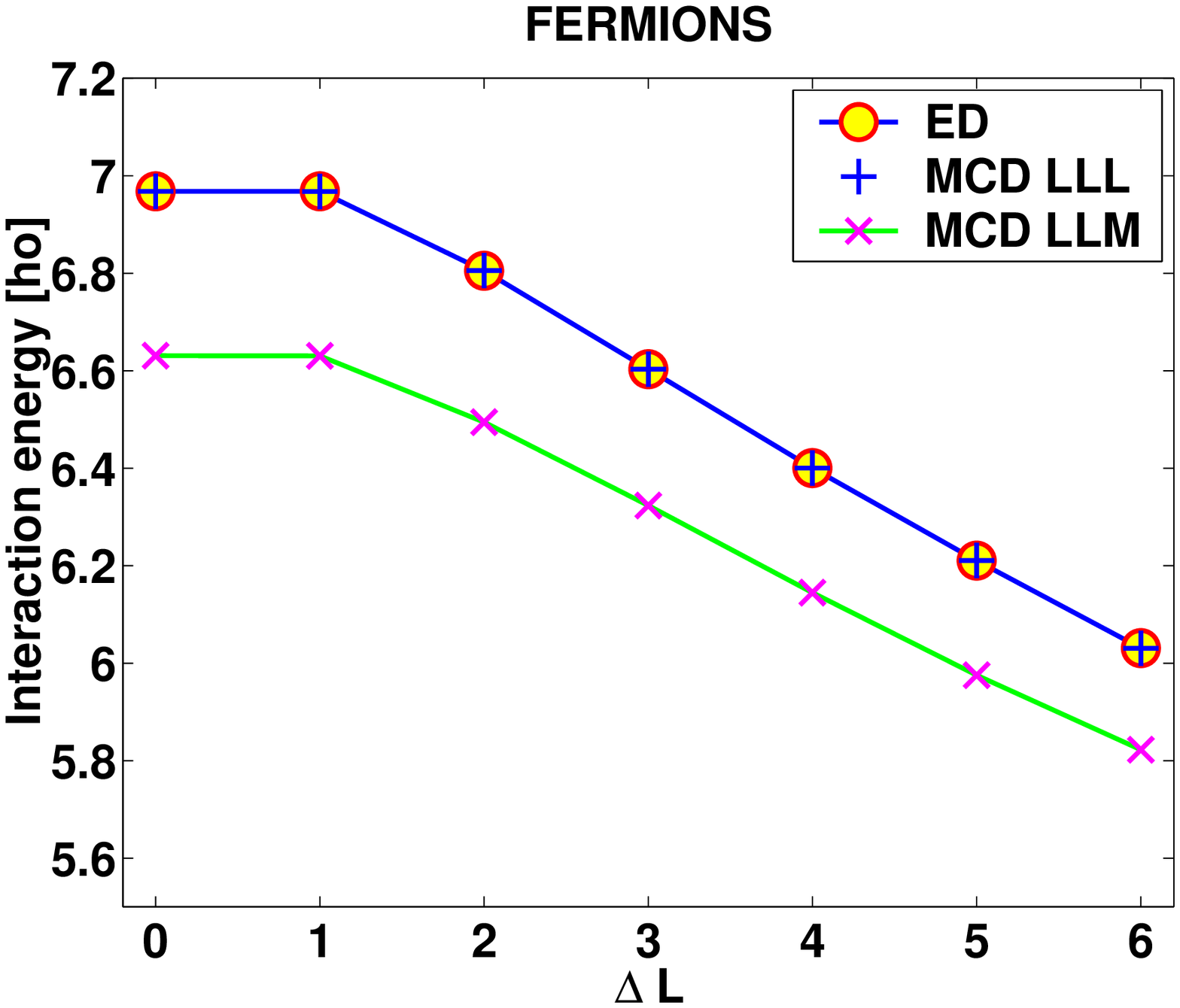}
  \includegraphics[width=0.49\columnwidth]{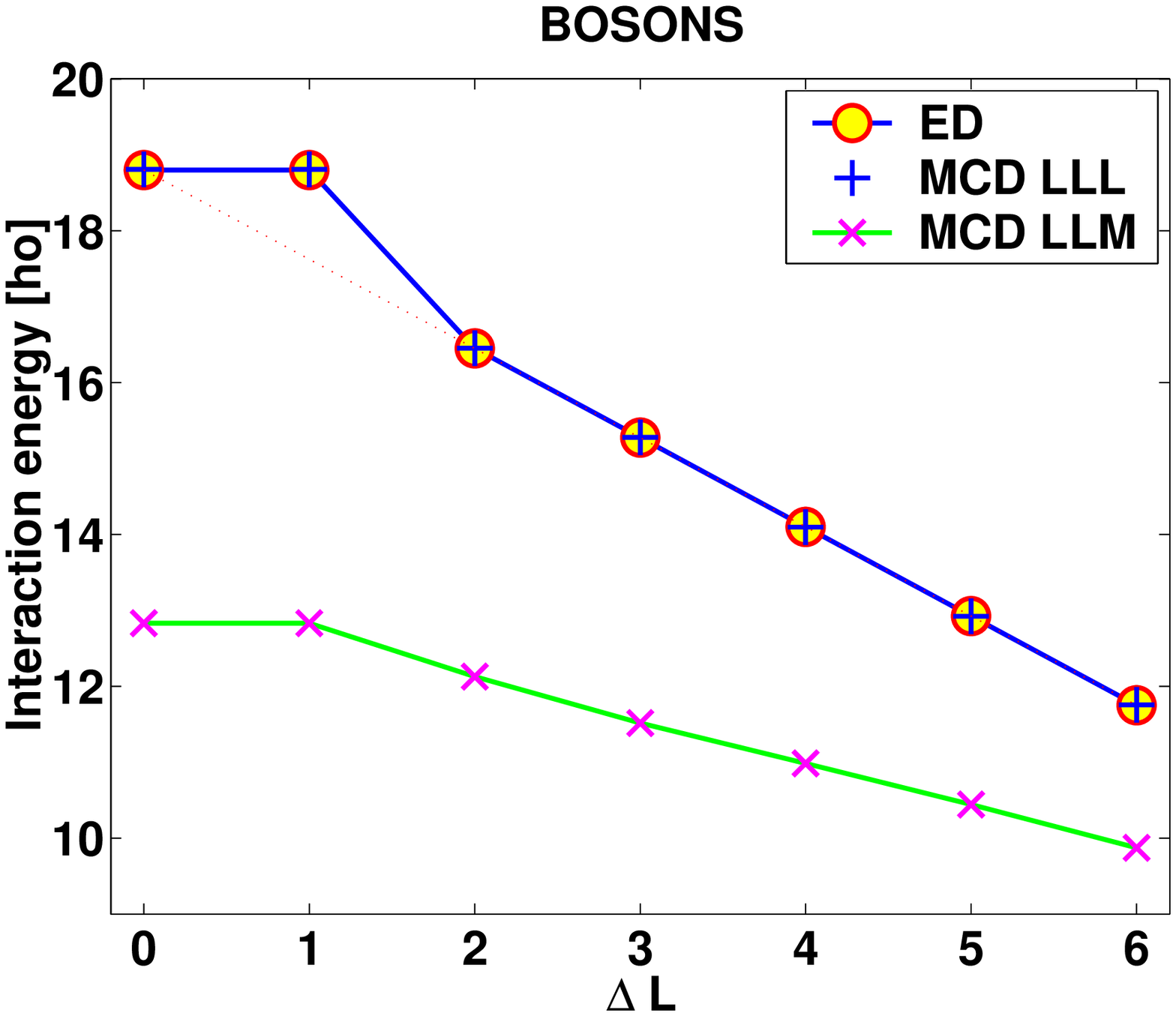}
\caption{Interaction energy (in units of $\hbar \omega$) for six
fermions and bosons as a function of the additional angular
momentum. For LLL, MCD is compared with the exact numbers, and perfect
agreement is found. The LLM energy shows the energy lowering by the
Jastrow factor.}
\label{MCD}
\end{figure}
As an example of the accuracy of the method, the interaction energies
of six fermions and bosons are shown in Fig.~\ref{MCD}. The data is
obtained from a few-minute runs, and the maximum error in the LLL is
0.0004 for fermions and 0.01 for bosons. The energy lowering obtained
by adding the Jastrow factor is significant. It is larger for bosons
because they lack the Pauli principle that keeps the spin-polarized
electrons of the fermion case further apart already on the LLL.

\begin{figure}[hbt]
\begin{center}
 \includegraphics[width=0.49\textwidth,angle=0]{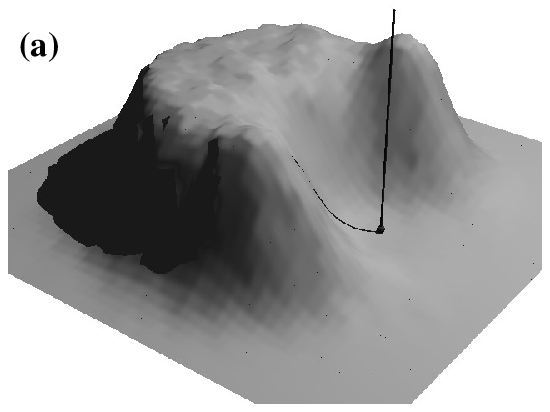}
 \includegraphics[width=0.49\textwidth,angle=0]{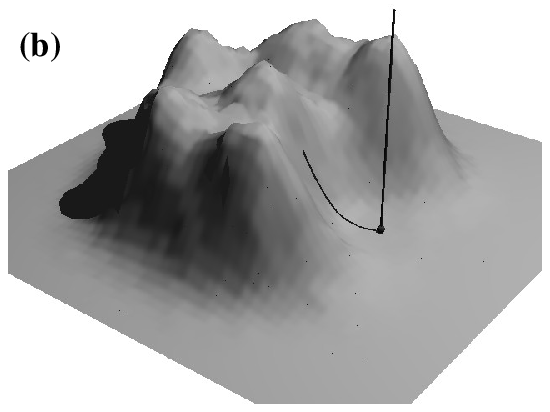}\\
\caption{ The charge density of a six-electron GaAs quantum dot, with
    a point-like impurity on the plane of the electrons (indicated by
    the line segment). The external confinement is $\hbar \omega =
    3$~meV and the magnetic field is 4~T in (a) and 6~T in (b).
    Particles start to localize at stronger magnetic field. }
\label{imp}
\end{center}
\end{figure}

A very interesting feature of the MCD method is that any classical
potential is easy to add. To demonstrate this, we have added a
point-charge impurity to the $N=6$ QD. The resulting electron density
is shown in Fig.~\ref{imp}.
One can see that the left side of the density in Fig.~\ref{imp}(a)
still looks flat like the one of MDD, but when the magnetic field gets
stronger, electrons start to localize, see Fig.~\ref{imp}(b). One can
clearly see the six peaks in density corresponding to the electrons in
the system. More details of MCD can be found from
Ref.~\onlinecite{SamiTkT}.

\begin{figure}[hbt]
  \includegraphics[width=0.49\columnwidth]{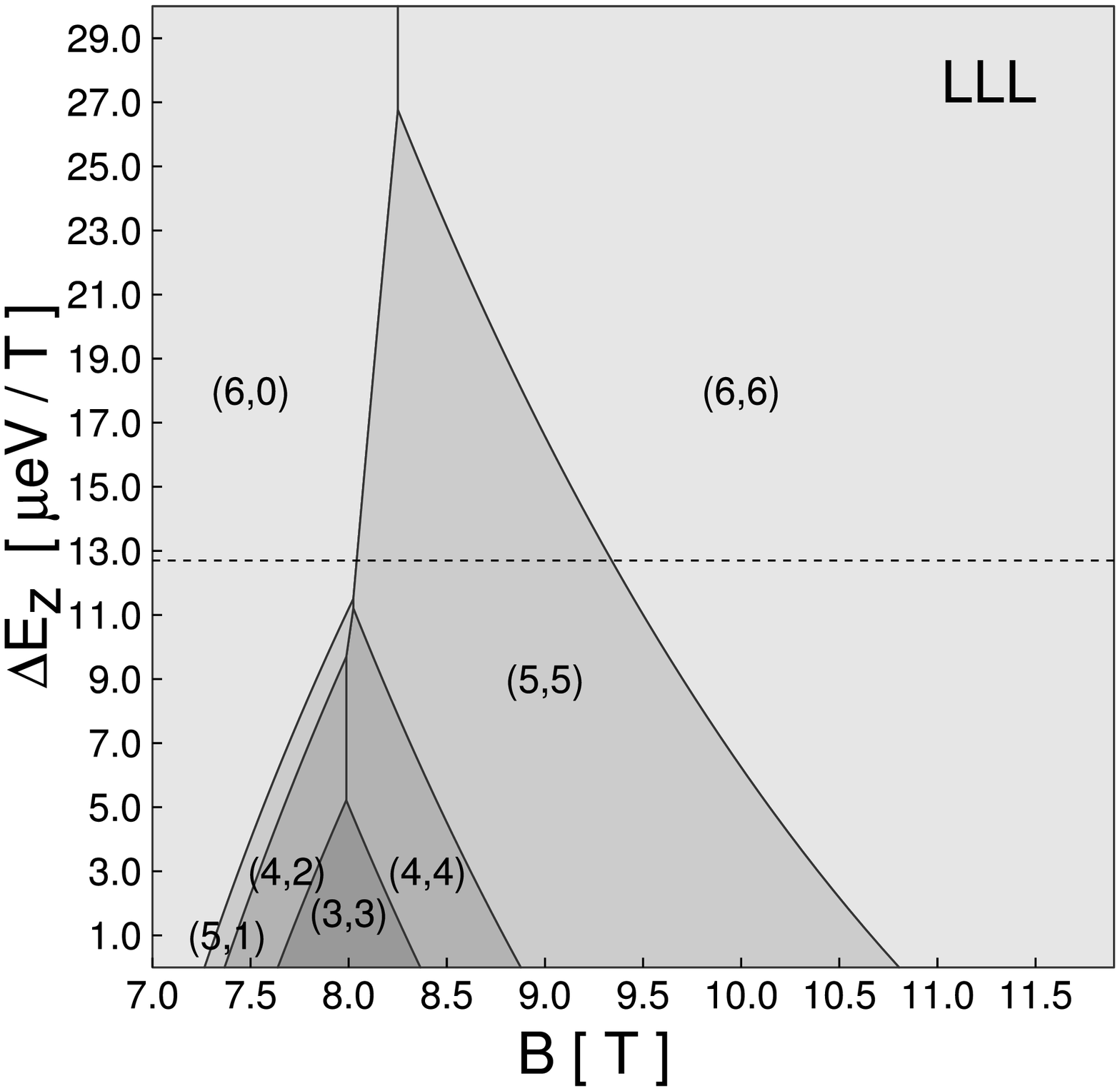}
  \includegraphics[width=0.49\columnwidth]{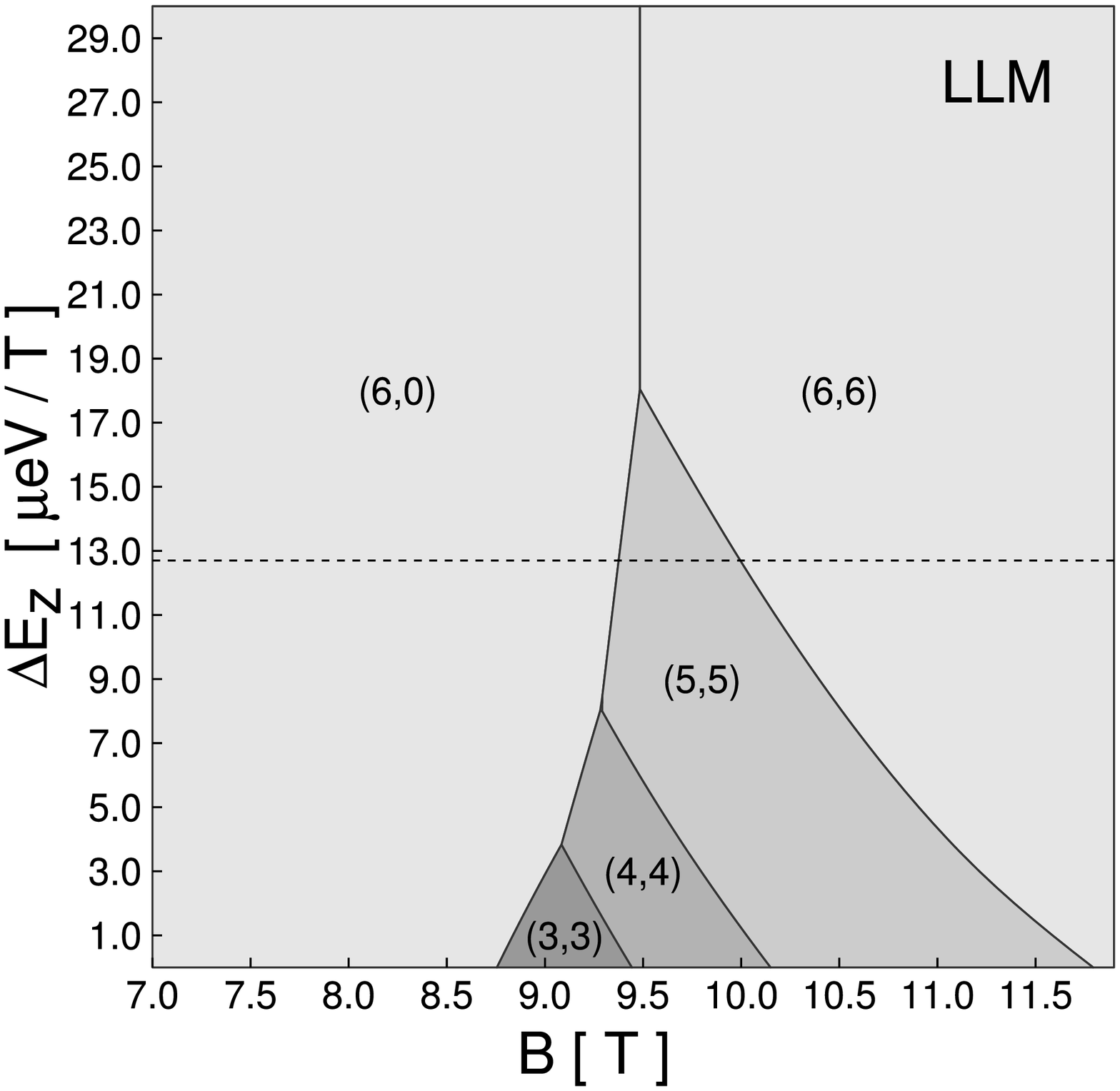}
\caption{Phase diagrams for six electrons in the lowest Landau level
approximation (left panel) and with Landau-level mixing (right panel).
The labeling of the states is: $N_{\uparrow}, \Delta L$, where
$N_{\uparrow}$ is the number of spin-up electrons and $\Delta L$ is
the additional angular momentum ($L_{\mathrm{MDD}}=15$ for six
electrons).  The vertical axis is the strength of the Zeeman coupling
per spin, $\Delta {E_z} = |\frac{1}{2} \mu_{\mathrm{B}} g^{\ast}|$,
the value of which in GaAs ($12.7\, \mathrm{\mu eV/T}$) is marked by
dashed lines in the figures.  Other parameters are: $m^{\ast}/m_0 =
0.067$, $\hbar \omega_{0} = 5\, \mathrm{meV}$ and
$\epsilon_{\mathrm{r}} = 12.4$.  The relative interaction strength $C$
varies from 1.23 ($B=7\, \mathrm{ T}$) to 1.01 ($B=12\, \mathrm{T}$).
}
\label{spin}
\end{figure}

Finally, there is also a possibility to find partially spin polarized
states after MDD\cite{RapidSpin}. Of course, this depends crucially on
the strength of the Zeeman coupling to spin. For realistic parameters
of the GaAs QD's, the $N=6$ QD has a ground state with one flipped
spin after the MDD state.  Here the LLL approximation has rather
different phase diagram of post-MDD states than the one with higher
levels also via the Jastrow factor, see Fig.~\ref{spin}.

\section{CONCLUSIONS}

We have constructed a variational wave function for quantum dots in
various magnetic field regimes.  The variational quantum Monte Carlo
method has shown to be accurate for all magnetic fields. The simple
single configuration Slater-Jastrow wave function can be used for many
cases, the possible exceptions being some of the states found in the
extreme magnetic fields where the system is in the fractional quantum
Hall effect regime. In this limit, we show that it is possible to
tailor a new type of a quantum Monte Carlo method that naturally
includes the multi-configurational nature of the system.

\section*{ACKNOWLEDGMENTS}

I would like to thank (in alphabetic order)
 Matti Alatalo,
 Bernardo Barbiellini,
 Veikko Halonen,
 Petteri Hyv\"onen,
 Karri Niemel\"a,
 Martti Puska,
 Stephanie Reimann,
 Henri Saarikoski,
 Sami Siljam\"aki,
 Juha Suorsa, and
 Viktor Sverdlov
for collaboration and 
 Sigridur Sif Gylfadottir,
 Meri Helle (n\'ee Marlo),
 Risto Nieminen, and
 Esa R\"as\"anen
also for their careful reading of the manuscript. This research is
supported by the Finnish Academy of Science and Letters and the
Academy of Finland, partly via its Centers of Excellence Program
(2000-2005).

\end{document}